\documentclass[]{agujournal}

\journalname{Earth's Future}
\newcommand{\about}{\mathord{\sim}}

\makeatletter
\def\@oddhead{}
\let\@evenhead\@oddhead
\makeatother

\begin{document}

%
%

\title{Evolving understanding of Antarctic ice-sheet physics and ambiguity in probabilistic sea-level projections}

%
%

\authors{Robert E. Kopp\affil{1}, Robert M. DeConto\affil{2}, Daniel A. Bader\affil{3}, Carling C. Hay\affil{1,4,8}, Radley M. Horton\affil{3}, Scott Kulp\affil{5}, Michael Oppenheimer\affil{6}, David Pollard\affil{7}, Benjamin H. Strauss\affil{5}}

\affiliation{1}{Department of Earth \& Planetary Sciences and Institute of Earth, Ocean \& Atmospheric Sciences, Rutgers University, New Brunswick, NJ, USA}
\affiliation{2}{Department of Geosciences, University of Massachusetts--Amherst, Amherst, MA, USA}
\affiliation{3}{Center for Climate Systems Research, Columbia University, New York, NY, USA}
\affiliation{4}{Department of Earth \& Planetary Sciences, Harvard University, Cambridge, MA, USA}
\affiliation{5}{Climate Central, Princeton, NJ, USA}
\affiliation{6}{Woodrow Wilson School of Public \& International Affairs and Department of Geosciences, Princeton University, Princeton, NJ, USA}
\affiliation{7}{Earth and Environmental Systems Institute, Pennsylvania State University, University Park, PA, USA}
\affiliation{8}{\emph{Current address:} Boston College, Boston, MA, USA}


\correspondingauthor{R. E. Kopp}{robert.kopp@rutgers.edu}


\begin{keypoints}
\item Incorporating ice-shelf hydrofracturing and ice-cliff collapse mechanisms highlights ambiguity in post-2050 sea-level projections.
\item These mechanisms make post-2050 sea level more heavily emissions dependent and can significantly revise high-emissions projections upwards.
\item Current Antarctic retreat by different processes than and exhibits little correlation with late-century changes.
\end{keypoints}

%
%

\begin{abstract}
Mechanisms such as ice-shelf hydrofracturing and ice-cliff collapse may rapidly increase discharge from marine-based ice sheets. Here, we link a probabilistic framework for sea-level projections to a small ensemble of Antarctic ice-sheet (AIS) simulations incorporating these physical processes to explore their influence on global-mean sea-level (GMSL) and relative sea-level (RSL). We compare the new projections to past results using expert assessment and structured expert elicitation about AIS changes. Under high greenhouse gas emissions (Representative Concentration Pathway [RCP] 8.5), median projected 21st century GMSL rise increases from 79 to 146 cm. Without protective measures, revised median RSL projections would by 2100 submerge land currently home to 153 million people, an increase of 44 million.  The use of a physical model, rather than simple parameterizations assuming constant acceleration of ice loss, increases forcing sensitivity: overlap between the central 90\% of simulations for 2100 for RCP 8.5 (93--243 cm) and RCP 2.6 (26--98 cm) is minimal. By 2300, the gap between median GMSL estimates for RCP 8.5 and RCP 2.6 reaches $>10$ m, with median RSL projections for RCP 8.5 jeopardizing land now occupied by 950 million people (vs. 167 million for RCP 2.6). The minimal correlation between the contribution of AIS to GMSL by 2050 and that in 2100 and beyond implies current sea-level observations cannot exclude future extreme outcomes. The sensitivity of post-2050 projections to deeply uncertain physics highlights the need for robust decision and adaptive management frameworks. 
\end{abstract}

%
%

\section{Introduction}
\label{sec:intro}

Probabilistic sea-level rise projections aim to characterize plausible Bayesian probability distributions -- usually conditional upon greenhouse gas emissions scenario -- of future global-mean sea-level (GMSL) and local relative sea-level (RSL) changes over time \citep[e.g.,][]{Kopp2014a,Kopp2016a,Mengel2016a,Slangen2014a,Jackson2016a,Wong2017a,Nauels2017a}. Those projections  explicitly labeled `probabilistic' generally aim to include estimates not just of central or `likely' ranges \citep[e.g.,][]{Church2013a}, but also estimates of the tails of probability distributions; others are conceptually similar but do not attempt to  estimate low-probability hazards \citep[e.g.,][]{Slangen2014a}. Many probabilistic projections are developed through frameworks that incorporate regional contributions to sea-level change, allowing them to be combined (for example) with estimated distributions of local flood frequencies to estimate an expected amplification of local flood frequencies over time \citep[e.g.][]{Buchanan2017a} and vertical elevation allowances needed to maintain expected flood frequency at a constant level \citep[e.g.,][]{Buchanan2016a,Slangen2017a}. There is a good degree of agreement on future GMSL among many of the studies producing probabilistic projections, as well as between these studies and the IPCC's Fifth Assessment Report \citep[AR5: ][]{Church2013a}. In some cases, this is by construction; for example, \citet{Kopp2014a} used ice-sheet projections based on a reconciliation of the structure expert elicitation study of \citet{Bamber2013a} with AR5. In other cases, this agreement represents independent lines of evidence leading to similar conclusions: for example, the agreement between AR5 and recent semi-empirical models relating GMSL change to global-mean temperature \citep{Kopp2016a,Mengel2016a}.

This agreement is not universal, however \citep[e.g.,][]{Jackson2016a}, and may be misleading. The response of polar ice sheets to forcing remains an area of `ambiguity' and `deep uncertainty' \citep{Kasperson2008a,Heal2014a}, for which it is currently impossible to identify a uniquely `correct' probability distribution. Approaches beyond historically calibrated statistical models \citep{Kopp2016a,Mengel2016a} and consensus-based expert assessment \citep{Church2013a} can provide additional reasonable ways of estimating probability distributions of the ice-sheet response. Notably, structured expert elicitation about the Antarctic ice sheet (AIS) response \citep{Bamber2013a} yielded a broader range than consensus-based AR5 expert assessment. Direct use of results from \citet{Bamber2013a}, without reconciliation with AR5, drives the higher projections in \citet{Jackson2016a}. Most physical models of the AIS response have been generally consistent with AR5 and \citet{Kopp2014a}: for example \citet{Ritz2015a}'s physical-statistical model estimated a 95th percentile AIS contribution to GMSL of  30 cm sea-level equivalent between 2000 and 2100, consistent with  \citet{Kopp2014a}'s  33-35 cm. \citet{Golledge2015a}'s deterministic model found a 39 cm contribution under high emissions and their higher, sub-grid interpolation of basal ice melt. However, \citet{DeConto2016a} (henceforth, DP16) found that the inclusion in a physical model of previously omitted  processes such as ice-shelf hydrofracturing and structural collapse of tall, marine-terminating ice cliffs has the potential to drive an order-of-magnitude increase in Antarctic mass-loss rates.

Ideally, the integration of process models into probabilistic frameworks such as those of \citet{Kopp2014a}  (henceforth, K14) and \citet{Jackson2016a} would involve the development and use of fast models -- or fast statistical emulators of more complex models -- in a mode that allows Monte Carlo sampling of key uncertainties and the conditioning of uncertain parameters on multiple observational lines of evidence. The development of such fast models or model emulators is an involved task, however, and the publication of DP16 triggered an increase in stakeholder interest and a demand for more expeditious approaches. For example, regional sea-level assessments for the City of Boston \citep{BRAG2016a} and State of California \citep{Cayan2016a} adopted the K14 framework but substituted, as a set of discrete samples, an ensemble of Antarctic ice-sheet projections from DP16. For the United States' Fourth National Climate Assessment, a U.S. Interagency Task Force on Sea-Level Rise report \citep{Sweet2017a} semi-quantitatively assessed how DP16's results might shift the probability distribution of future GMSL change.

Here, we extend the approach of \citet{BRAG2016a} and \citet{Cayan2016a} to a global scale, substituting  DP16's AIS ensembles for  K14's expert assessment- and expert elicitation-based probability distribution. This substitution allows more complex temporal dynamics than the simple assumptions of constant acceleration that underlie many expert-judgement-based projections \citep[e.g.,][]{Bamber2013a,Little2013a,Little2013b,Kopp2014a}. It also allows identification of the importance of different physical assumptions regarding AIS for total GMSL and RSL projections. However, this approach comes with some limitations. DP16 did not originally develop their ensemble to produce a probability distribution. Instead, they sampled their key physical parameters from a discrete, somewhat arbitrary set of values, and they integrated paleo-observations via a simple pass/fail test. A more probabilistic approach would have employed continuous prior probability distributions for key parameters and integrated observations -- potentially instrumental observations as well as paleo-constraints -- via Bayesian updating. The more limited approach taken means their ensemble may give excessive weight to certain ensemble members, while also having unrealistically thin tails delimited by the discrete values selected for the prior parameter values. 

This substitution approach differs from that taken by \citet{Le-Bars2017a}, who sought to integrate DP16 projections into comprehensive projections of GMSL rise between 2000 and 2100. They fit a normal distribution to DP16 ensembles for 2100, while also introducing a linear dependence of AIS mass loss on global mean surface temperature. In contrast to the approach here, their approach neglects the non-normality in the DP16 ensemble and gives significant weight to values represented in the tails of their fitted normal distributions but not within the physically modeled DP16 projection ensemble. The approach in their current paper is thus more conservative with respect to extrapolation beyond the DP16 ensemble values. In addition, by using the DP16 projections directly, this study  considers not just a single time point but also the course of sea level over time. Leveraging the K14 projection framework, this study  also ties global projections to their regional manifestations.

Because of the limitations associated with using the DP16 projections directly, the resulting GMSL and RSL projections should not be viewed as constituting well-constructed Bayesian probability distributions. Accordingly, we refer to the resulting distributions as \textit{simulation frequency distributions}, not \textit{probability distributions}; this terminological choice parallels that of \citet{NAS2016a} in the context of social cost of carbon dioxide estimates. These simulation frequency distributions can be used in some contexts -- e.g., decision frameworks leveraging multiple alternative probability distributions -- as if they were probability distributions. Doing so effectively treats the DP16 choices of parameter values as though they constituted a well-constructed, equally weighted prior, and the DP16 paleo-constraints as though they were well-represented by uniform likelihood distributions. Given the weaknesses in these assumptions, we would advise against using these simulation frequency distributions in isolation as new `best-estimate' probability distributions. 

\section{Methods}

\subsection{Projections Framework}

The framework  employed to generate GMSL and RSL projections in this analysis is based on that of K14.  The K14 framework combines multiple lines of information to construct probability distributions for key contributors to GMSL and RSL change. It employs a joint probability distribution for global mean thermal expansion and regional ocean dynamics derived from the Coupled Model Intercomparison Project Phase 5 (CMIP5) \citep{Taylor2012a} ensemble. Its projections of glacier mass-balance changes are derived from the \citet{Marzeion2012a} surface mass-balance model, forced by the CMIP5 ensemble. Following the approach of \citet{Rahmstorf2012a}, its projections of the global-mean contribution of anthropogenic changes in land-water storage are based upon historical relationships between human population, dam construction, and groundwater withdrawal \citep{Chao2008a,Wada2012a,Konikow2011a}. The regional contributions of non-climatic effects such as glacio-isostatic adjustment, tectonics, and sediment compaction are based upon a spatiotemporal  statistical model of tide-gauge observations. Ice sheet contributions are derived from the AR5 expert assessment and the structured expert elicitation of \citet{Bamber2013a}, as described below. Glacier and ice sheet projections are translated into RSL changes using static-equilibrium fingerprints for eighteen glacier regions, the Greenland Ice Sheet, the West Antarctic Ice Sheet (WAIS), and the East Antarctic Ice Sheet (EAIS) \citep{Mitrovica2011a}.

\citet{Bamber2013a} elicited from fourteen experts central 90\% probability estimates for the rate of GMSL rise in 2100 due to the Greenland ice sheet, WAIS and EAIS. They did not distinguish between surface-mass balance and dynamic contributions, nor did they distinguish between emissions scenarios. In turning these rates into cumulative 21st century GMSL rise contributions, they assumed a linear increase in rates based on the experts' rate estimates for the last decade and for 2100. 

AR5 assessed the likely (central 66\% probability; see exegesis by \citet{Church2013b}) range of Greenland and Antarctic contributions in 2080-2099, distinguishing between surface-mass balance and dynamic terms. They did not distinguish between EAIS and WAIS. For the dynamic AIS contribution, they did not distinguish among RCPs.

K14 combined the \citet{Bamber2013a} and AR5 approaches in a manner intended to retain consistency with the likely ranges of AR5. In particular, K14: (1) calculated probability distributions for EAIS, WAIS, and Greenland changes over time from \citet{Bamber2013a}, assuming linear changes in rates; (2) calculated probability distributions for AIS and Greenland over time based on the AR5 likely ranges for 2080--2099, again assuming linear changes in rates to achieve these values; (3) added a time-varying factor to the first set of distributions so the medians of the two sets align; (4) separated the AR5-derived Antarctic distribution into EAIS and WAIS terms by assuming the EAIS/WAIS ratio is the same as in the median projection from the first set; and (5) applied multipliers (separately for values greater than and less than the median) to the difference of the values in the final distribution from the distribution's median, so that the central 66\% probability range matches that of AR5.

\subsection{Revised Antarctic projections}

In this paper, we compare two sets of projections. The first, which we label K14, follows the original methodology of K14, extended in space and time. The second, which we label DP16, replaces the AIS projections of K14 with projections based on new physical modeling \citep{DeConto2016a}. These processes include the influence of surface meltwater, driven by summer temperatures above freezing and the increasing ratio of rain to snow in a warming climate, on the penetration into ice shelves of surface crevasses that can lead to hydrofracturing. Hence, in DP16, buttressing ice shelves can thin or be lost entirely due to sub-ice ocean warming, the extensive spread of surface meltwater, or a combination of the two. In places where thick, marine-terminating grounding lines have lost their buttressing ice shelves, a wastage rate of ice is applied locally at the tidewater grounding line in places where vertical ice cliffs are tall enough to produce stresses that exceed the yield strength of the ice (see \citet{DeConto2016a} and \citet{Pollard2015a} for complete formulation).

Three uncertain but key model parameters relate to (1) the rate of sub-ice shelf melt rates in response to warming ocean temperatures (OCFAC), (2) the sensitivity of crevasse penetration to meltwater input (hydrofracturing) (CREVLIQ), and (3) the maximum rate of cliff collapse (VCLIF). Because, as discussed below, there are no modern analogues to widespread ice-cliff failure, model performance cannot be adequately judged relative to Holocene or recent trends in ice-sheet behavior. Instead, the new model physics were tested relative to past episodes of ice sheet retreat during the Pliocene ($\about 3$ Ma) and the Last Interglacial (LIG, $\about 125$ Ma), when Antarctic ocean and surface air temperatures were warmer than today \citep{Capron2014a,Rovere2014a}. The three key parameters were varied systematically. From an initial 64 versions of the ice sheet model, 29 were found to satisfy both Pliocene and LIG sea-level targets, with Antarctic contributions to GMSL ranging between 5 to 15 m (Pliocene) and 3.6 to 7.4 m (LIG). The range of oceanic melt rate model parameters passing the Pliocene and LIG sea-level tests are comparable to those determined from a large, 625-member ensemble of the last deglacial retreat of the West Antarctic Ice Sheet using the same ice sheet model \citep{Pollard2016a}; however, the deglacial simulations do not provide guidance on hydrofracturing and ice-cliff physics, because the background climate was too cold to trigger those processes. 

One challenge of formulating a parameterization of ice-cliff physics is the lack of observations of marine-terminating ice without buttressing ice shelves and of sufficient thickness ($\about 1000$ m) to allow subaerial ice cliffs tall enough ($\about 100$ m) to drive structural collapse  \citep{Bassis2011a}. The few calving fronts of this scale that exist today (e.g., Helheim and Jakobshavn Glaciers on Greenland, and Crane Glacier on the Antarctic Peninsula) are experiencing rates of calving and structural failure at the terminus, comparable to the seaward flow of the glaciers, on the order of $\about 2$ to $>12$ km/yr \citep[e.g.,][]{Howat2008a,Joughin2014a,Wuite2015a}. Unlike several major Antarctic outlet glaciers, these Greenland outlet glaciers are in relatively narrow (5--12 km wide), restricted fjords, with substantial m\'elange (a mix of ice bergs and sea ice that can provide some supporting buttressing/back pressure at the terminus), and supportive, lateral shear along the fjord walls. Hence, using observed rates of cliff collapse to constrain the model physics representing these processes could lead to underestimates.

In Antarctica, there is potential for much wider ice cliffs to form along vast stretches of the coastline if floating ice tongues and shelves are lost. For example, the throat of Thwaites Glacier is about 120 km wide, but at present its grounding line is mostly on bedrock too shallow (about 600 m deep; \citet{Millan2017a}) to drive extensive structural failure at the terminus \citep{Bassis2011a}.  In the DP16, the highest maximum allowable rate of cliff collapse (VCLIF) -- the maximum horizontal rate of ice loss applied at the marine "tidewater'' calving terminus where ice cliffs are tall enough to generate stresses that exceed the strength of the ice -- is 5 km/yr. This rate is about half the rate of mass wastage at the front of Jakobshavn, which currently has a relatively stable terminus position but is flowing seaward at $> 12$ km/year \citep{Joughin2012a}. To include the potential for even faster rates of ice sheet mass loss than in the existing model formulation, future work should consider a wider range of parameter space.

We note that the paleo-sea-level targets used to test and calibrate the model physics provide limited guidance regarding potential rates of ice-sheet retreat. While \citet{Kopp2009a} do provide an estimate of the rate of sea-level rise contributed by the Antarctic Ice Sheet during LIG retreat, both their temporal resolution and their ability to attribute GMSL changes to AIS are limited. Moreover, given limited Antarctic atmospheric warming during the LIG relative to the Pliocene, initial WAIS retreat was more likely driven by oceanic warming than atmospheric warming \citep{DeConto2016a}, and therefore offers little in terms of validating rates of retreat driven by extensive surface melt, hydrofracture, and cliff collapse.

As described in \citet{DeConto2016a}, the 29 versions of the ice sheet model satisfying  geological constraints were used to simulate future ice-sheet retreat following RCP 2.6, 4.5, and 8.5 greenhouse gas pathways. In the future simulations, time-evolving oceanic melt rates were driven by NCAR CCSM4 \citep{Gent2011a,Shields2016a} sub-surface ocean temperatures. Surface mass balance and meltwater production rates were calculated from monthly air temperatures and precipitation provided by the RegCM3 regional climate model \citep{Pal2007a} run offline and bias-corrected relative to a modern climatology \citep{DeConto2016a,LeBrocq2010a}. 

Coupled atmosphere-ocean models are known to struggle with subsurface ocean temperatures in the circum-Antarctic \citep{Little2016a}. To minimize the effects of a general cold bias in NCAR CCSM4 Antarctic Shelf Bottom Water in the Amundsen and Bellingshausen Seas, a correction of 3$^\circ$C was applied to ocean temperatures at 400 m depth. This bias correction is meant to compensate for the recent warming observed there  \citep{Schmidtko2014a}. The correction is greater than the actual temperature offset, but given the formulation of the sub-ice melt rate parameterization used in DP16, a 3$^\circ$C correction is required to bring modern oceanic sub-ice shelf melt rates closer to observations \citep{Rignot2013a}. The effect of not using the ocean temperature/melt-rate correction in future simulations is shown in Supporting Information. 

\subsection{Detection simulation}

To simulate the process by which new observations of GMSL change can help detect whether the world is on a path leading to high or low levels of GMSL rise, we first define GMSL scenarios in a manner similar to \citet{Sweet2017a}. In particular, we pool the simulations of GMSL rise under RCP 2.6, RCP 4.5, and RCP 8.5, and then filter the pooled set to arrive at sets of simulations consistent with either $50 \pm 10$ cm or $200 \pm 10$ cm of GMSL rise between 2000 and 2100. We use the 5th-95th percentile range of these filtered sets to define scenario time paths. At each decade from 2000 to 2100, for each scenario, we compute the simulation frequency distribution of GMSL rise in 2100, conditional upon the observed GMSL in the decade being within the bounds of the scenario's time path. Finally, we compare the resulting conditional distributions to assess the detectability in a given decade of the difference between a pathway leading to about 50 cm of GMSL rise in 2100 and one leading to about 200 cm of GMSL rise.

\subsection{Extensions of the spatial and temporal domain}

The projections framework in this paper has a more extended spatial domain than the original K14 projections. While the original K14 projections were generated only at the precise location  of tide gauges, here we also generate projections at points on a $2^\circ \times 2^\circ$-resolution global grid that intersect world coastlines. At these points, we use the spatiotemporal statistical model described in K14 to estimate (with larger errors than at the tide-gauge sites) the long-term, non-climatic, background contribution to RSL change. The projection assumes that the background rate of change estimated from tide-gauge data continues unchanged over the duration of the projections. 

The projection framework also has a more extended temporal domain than the original K14 projections. Whereas the original K14 projections end in 2200, here we generated projections to 2300. This extension requires no computational modifications to the K14 framework. However, we regard this time frame as more appropriate when considering projections in which Antarctic ice sheet behavior are determined by a physical model rather than by a simple, temporally quadratic projection.

\subsection{Assessment of population exposure}

As an integrative metric of RSL changes, we assess the population currently occupying land threatened with submergence under different sea level rise projections. To do this, we compare land elevations from NASA's 1-arcsec SRTM 3.0 digital elevation model \citep{NASA-JPL2013a} against nearest-neighbor water elevations derived from adding the K14 and DP16 projection grids to measured local mean sea surface elevation augmented by a modeled tidal supplement.  We intersect the resulting inundation surfaces with contemporary population \citep{Bright2011a} and national boundary \citep{GADM2012a} data to estimate current national populations occupying land at risk of permanent submergence. SRTM data are the most practical option and widely used for global coastal exposure research, but bias estimates low \citep{Kulp2016a}. For each set of sea-level rise projections, we assess the population exposed assuming each grid cell followed its 50th, 5th, or 95th percentile RSL projection. Further details are provided in the Supporting Information.  We emphasize that the resulting values are not projections of the impacts of RSL change, which would require a dynamic model considering both population growth and migration away from inundated regions; rather, population here serves as a convenient integrative metric.

\section{Results}

\begin{figure}[tb]
\centering
\includegraphics[width=20pc]{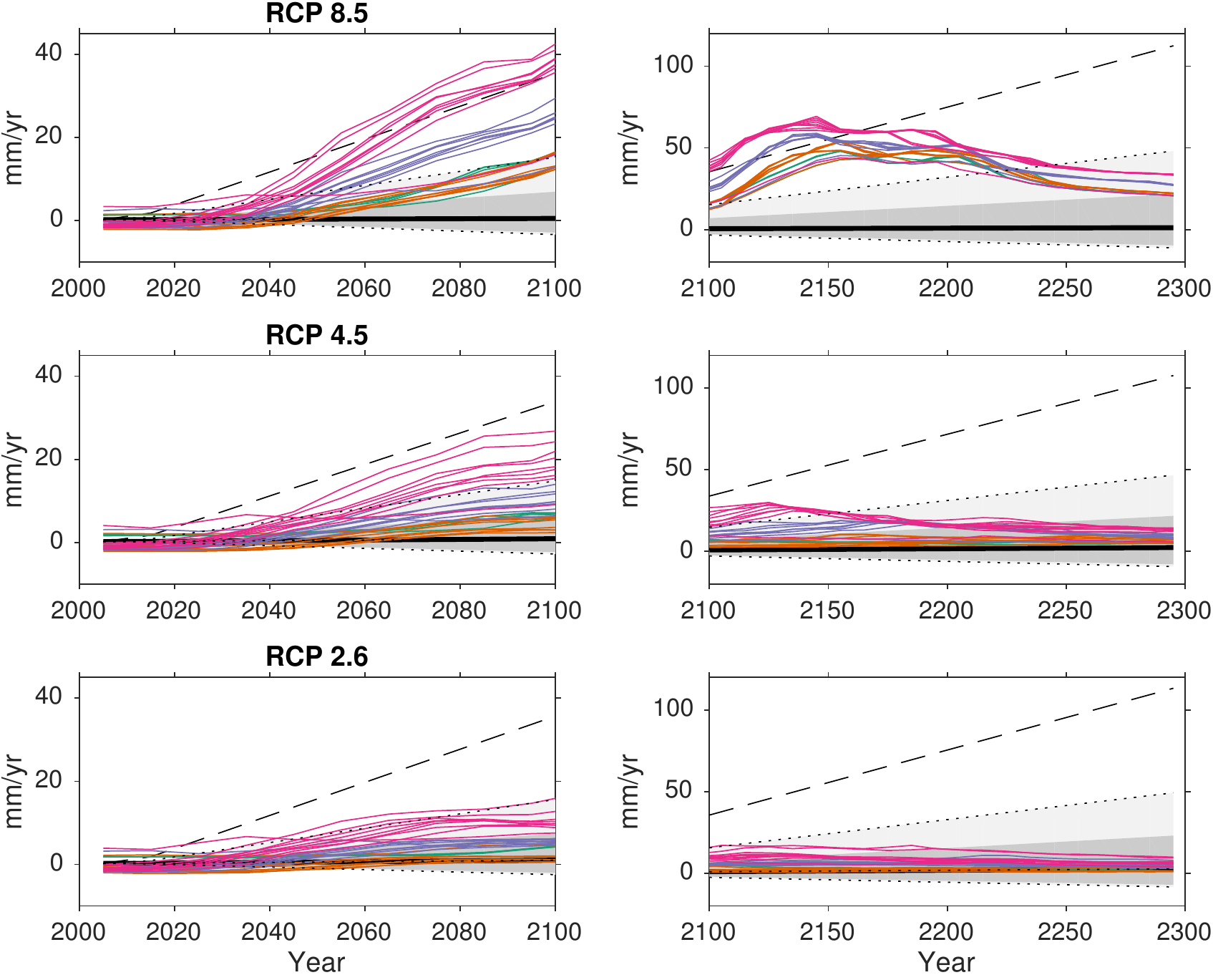}
\caption{Rates of contribution of the Antarctic ice sheet to GMSL under the three RCPs. Dark/light shaded areas represent 5--95th and 0.5th--99.5th percentile of K14. Dashed black line represents 99.9th percentile of K14. Colored curves are DP16 runs, with colors  reflecting different maximum rates of ice-cliff collapse [VCLIF] (green: no ice cliff collapse; orange: 1 km/yr; purple: 3 km/yr; magenta: 5 km/yr). Left panels show 2000--2100, right panels show 2100--2300. Note change of horizontal and vertical scales.  }
\label{fig:rate-AISoldandnew}
\end{figure}

\begin{figure}[tb]
\centering
\includegraphics[width=20pc]{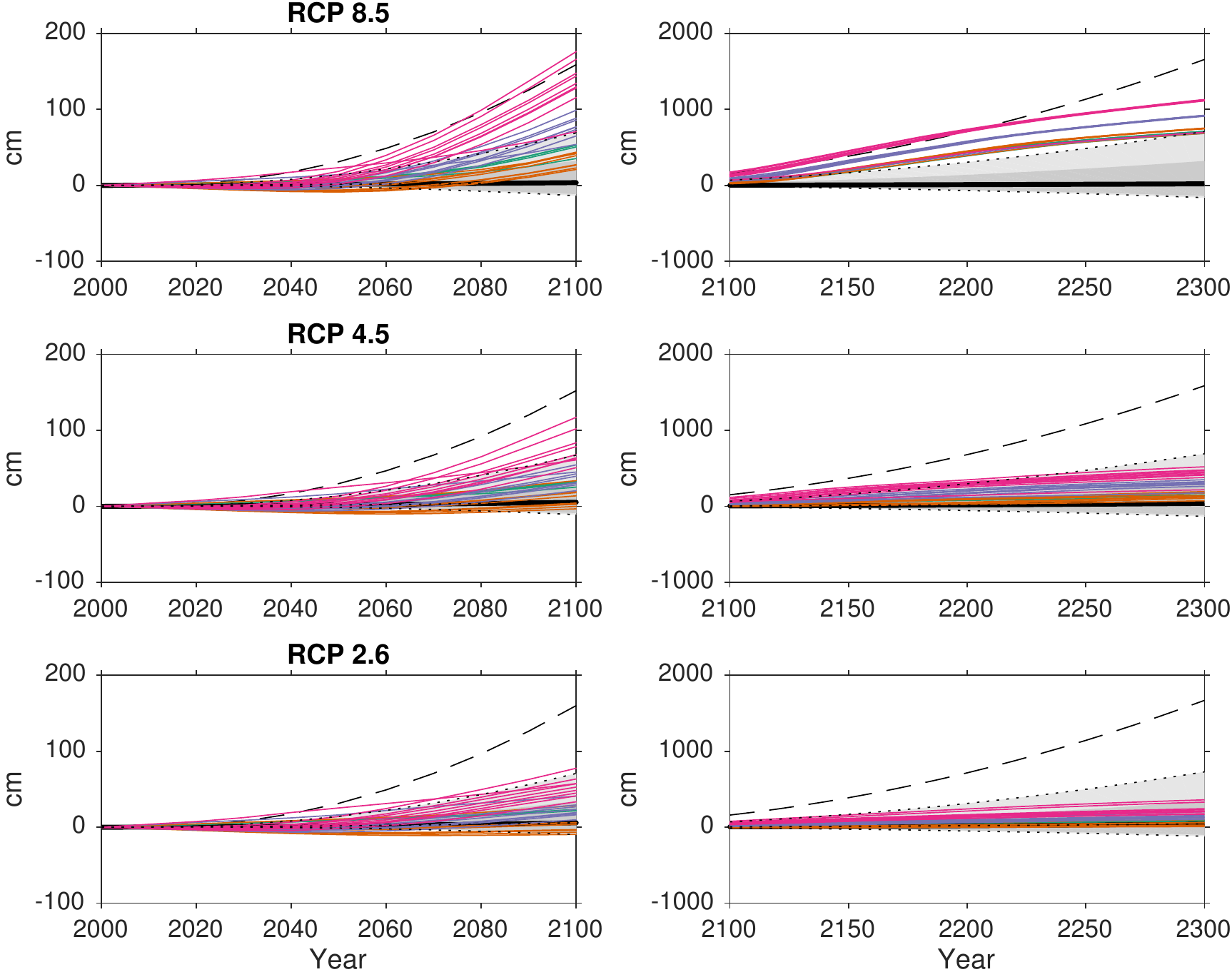}
\caption{Projections of the Antarctic ice sheet contribution to GMSL under the three RCPs. Dark/light shaded areas represent 5--95th and 0.5th--99.5th percentile of K14. Dotted black line represents 99.9th percentile of K14. Colored curves are DP16 runs, with colors  reflecting different maximum rates of ice-cliff collapse (green: no ice cliff collapse; orange: 1 km/yr; purple: 3 km/yr; magenta: 5 km/yr). Left panels show 2000--2100, right panels show 2100--2300. Note change of horizontal and vertical scales. }
\label{fig:AISoldandnew}
\end{figure}

The K14 Antarctic projections -- like those of \citet{Bamber2013a} and \citet{Little2013a,Little2013b}, among others -- assumed that changes in the rates of change in ice-sheet  mass balance occurred linearly. For example, \citet{Bamber2013a} elicited expert opinion on the rate of Antarctic ice sheet mass change in 2100, and assumed that the elicited rate was achieved following a linear growth rate. The result is a quadratic change in ice volume over time. K14 took the same approach (Figures \ref{fig:rate-AISoldandnew} and \ref{fig:AISoldandnew}). By contrast, process modeling as in DP16 shows considerably more complex behavior, with periods of rapid increases in mass loss rate as individual sectors of ice sheet collapse, and other intervals of stable or declining rates of retreat (Figure \ref{fig:rate-AISoldandnew}). Sizable non-linearity appears in all simulations under strong forcing (Figure \ref{fig:rate-AISoldandnew}, RCP 8.5) and under all forcings in almost all simulations with high maximum rates of ice-cliff collapse (Figure \ref{fig:rate-AISoldandnew}, purple and magenta curves).

In the first half of the 21st century, the range spanned by DP16 Antarctic projections  is similar to that spanned by K14 (-10 to +23 cm contribution to GMSL  in 2050, vs. a 1st--99th percentile range of -2 to +14 cm under K14).  Both sets of projections show minimal emissions-scenario dependency in the first half of the 21st century. The central tendency among the DP16 projections is slightly higher, with a median contribution to GMSL of about +5 cm under DP16, compared to a median of +2 cm under K14.  These slightly higher values are driven by the ocean-temperature bias correction, which is needed to improve consistency with observed oceanic sub-ice melt rates in the Amundsen and Bellingshausen Sea sectors of West Antarctica \citep{Rignot2013a}. Without this  correction, there is a tendency toward Antarctic growth in the early decades of the century (in 2050, RCP 8.5: median -3 cm, range of -9 to +12 cm, RCP 2.6: median -2 cm, range of -10 to +6 cm). However, even with the bias correction, Antarctica's median contribution to GMSL is still ~0.1 mm/yr, which is about a factor of 3 less than that currently observed for the early 21st-century \citep{Shepherd2012a,Church2013a,Harig2015a}.  Overall, the substitution of DP16 has a very limited effect on mid-century GMSL projections (Table \ref{tab:GMSL}). 

Under strong forcing, the overall picture changes dramatically  by the end of the 21st century, with several of the DP16 simulations leading to AIS contributions to GMSL exceeding +1 m by 2100  under RCP 8.5 (Figure \ref{fig:AISoldandnew}).  These high projections are driven by high maximum rates of ice-cliff retreat (VCIF = 5 km/yr) in combination with non-zero sensitivity of ice shelves to hydrofracturing (CREVLIQ > 0) (Figures \ref{fig:AISoldandnew}, \ref{Sfig:AISoldandnew-CREVLIQ}). As a consequence, the median DP16 GMSL projections for 2100  under RCP 8.5 reaches 146 cm, the 98th percentile projection under K14. The low tail is curtailed by the incorporation of physical modeling, with a 1st percentile of 80 cm exceeding the median of K14. With a high VCLIF, the median GMSL projection reaches 213 cm (in excess of the 99th percentile of K14); with no cliff collapse mechanism or no hydrofracturing, it is reduced to about 125 cm (96th percentile of K14) (Supporting Table S3, S4). 

DP16 RSL projections indicate the risk of significant changes to the global coastline by 2100. Without protective measures, the 5th--95th percentiles of RSL projections under DP16 would inundate land currently home to 106--236 million people. This contrasts with 82--154 million people under the K14 projections (Table \ref{tab:totalpopulationexposure}, full table in Data Set 6). 

A significant enhancement of the AIS contribution to GMSL also occurs for 2100 under moderate forcing: the median DP16 total GMSL projection of 91 cm under RCP 4.5 is consistent with the 93rd percentile of K14. The low tail is modestly curtailed: the DP16 1st--99th percentile values for RCP 4.5 (39--180 cm) resemble the K14 9th--99.8th percentile range. Under low forcing (RCP 2.6), there is little effect, with the DP16 1st--99th percentile range (18--111 cm) resembling the K14 0.5th--99th percentile range.

These differences build over the 22nd and 23rd century. By 2300, under RCP 8.5, the median DP16 GMSL projection of 11.7 m exceeds the K14 99th percentile. Although the ice-cliff collapse mechanism contributes to this projection, the median projection remains as high as 10.0 m by 2300 even without it  (Supporting Table S3). Without protective measures, median DP16 RSL projections would submerge land currently home to 950 million people worldwide, a roughly three-fold increase relative to K14 (Table \ref{tab:totalpopulationexposure}). The DP16 1st--99th percentile range  (8.6--17.5 m) resembles the K14 97th--99.8th percentile range. Under RCP 4.5, the median DP16 GMSL projection of 4.2 m resembles the K14 90th percentile, and the DP16 1st--99th percentile range (1.6--8.1 m) resembles the  K14 42nd--98th percentile range. The median is reduced to 3.0 m (75th percentile of K14) without the ice-cliff collapse mechanism, and 3.2 m (79th percentile of K14) without the hydrofracturing mechanism. Under RCP 2.6, by contrast, the median DP16 GMSL projection of 1.4 m  matches the K14 median, and the DP16 1st--99th percentile range (0.2--4.0 m) resembles the K14 14th--92nd percentile range.

\begin{figure}[tb]
\centering
\includegraphics[width=20pc]{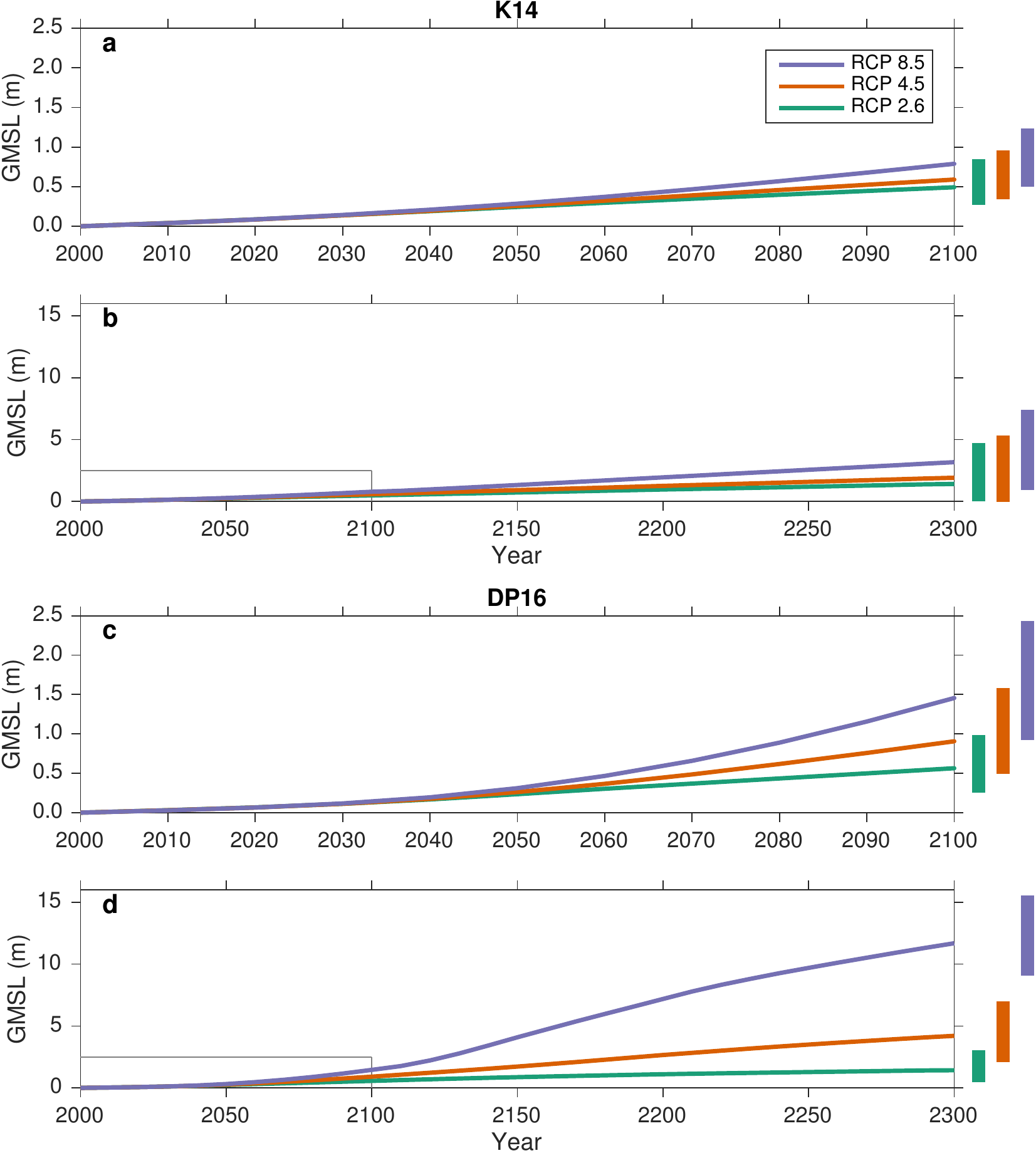}
\caption{Projections of GMSL rise for the three RCPs under K14 (a-b) and DP16 (c-d). Lines indicate median; boxes indicate 5th--95th percentile range for 2100 (a,c) and 2300 (b,d). Light grey lines in (b) and (d) indicate axes limits of (a) and (c). }
\label{fig:GMSLoldandnew}
\end{figure}

Taken together, the incorporation of the DP16 AIS ensemble pulls the projections much higher by 2100 and beyond under RCP 4.5 and especially RCP 8.5 (Figure \ref{fig:GMSLoldandnew}).  It thus leads to a significant reduction in overlap among projections  of GMSL change based upon different emissions scenarios. This is to be expected based on the difference in construction between the K14 Antarctic projections and the DP16 projections. In K14, as in AR5, AIS surface mass balance is scenario-dependent, but the ice-sheet dynamic term is treated as scenario-independent: it is assumed that the uncertainty in physical understanding of ice-sheet behavior swamps the forcing uncertainty. By contrast, the physical model of DP16 yields a strong forcing dependence.

\begin{table}[tb]
\caption{Projections of GMSL rise (cm)}
\centering
\scriptsize
\begin{tabular}{l l l l l l}
  & 50 & 17--83 & 5--95 & 1--99 & 99.9 \\
\hline
K14 \\
RCP 8.5 \\
2050 & 29 & 24--34 & 21--39 & 17--46 & 59 \\
2100 & 79 & 62--101 & 51--123 & 40--159 & 232 \\
2200 & 195 & 131--284 & 94--380 & 64--552 & 886 \\
2300 & 318 & 175--516 & 98--737 & 37--1093 & 1929 \\
RCP 4.5 \\
2050 & 26 & 21--31 & 18--35 & 15--41 & 55 \\
2100 & 59 & 44--77 & 35--95 & 26--128 & 205 \\
2200 & 126 & 70--197 & 36--278 & 8--433 & 780 \\
2300 & 192 & 70--349 & 0--531 & -55--900 & 1717 \\
RCP 2.6 \\
2050 & 24 & 20--29 & 18--33 & 15--40 & 55 \\
2100 & 49 & 36--66 & 28--84 & 20--120 & 203 \\
2200 & 97 & 48--163 & 23--242 & 6--406 & 803 \\
2300 & 142 & 32--288 & -22--470 & -57--847 & 1773 \\
\hline
DP16 \\
RCP 8.5 \\
2050 & 31 & 22--40 & 17--48 & 13--54 & 59 \\
2100 & 146 & 109--209 & 93--243 & 80--267 & 297 \\
2200 & 719 & 595--896 & 558--962 & 525--1049 & 1193 \\
2300 & 1169 & 980--1409 & 913--1552 & 861--1751 & 2006 \\
RCP 4.5 \\
2050 & 26 & 18--36 & 14--43 & 10--52 & 57 \\
2100 & 91 & 66--125 & 50--158 & 39--180 & 197 \\
2200 & 266 & 176--396 & 133--455 & 102--510 & 594 \\
2300 & 421 & 275--595 & 211--696 & 163--806 & 995 \\
RCP 2.6 \\
2050 & 23 & 16--33 & 12--41 & 9--50 & 54 \\
2100 & 56 & 37--78 & 26--98 & 18--111 & 122 \\
2200 & 110 & 70--161 & 47--206 & 30--250 & 314 \\
2300 & 142 & 83--230 & 50--300 & 22--404 & 552 \\
\hline
\multicolumn5l{Columns indicate percentiles of simulation frequency distributions.}
\end{tabular}
\label{tab:GMSL}
\end{table}

\begin{table*}[tb]
\caption{Population exposure (millions of people)}
\centering
\scriptsize
\begin{tabular}{cccccc}
\multicolumn{6}{l}{{Current population occupying land exposed to inundation under 2100 RSL projections}}   \\\hline
Region         & Total Pop.            & RCP 2.6/K14        & RCP 2.6/DP16       & RCP 8.5/K14        & RCP 8.5/DP16        \\ \hline
World      & 6,836 & 94.3 (73.3 - 127.6) & 97.4 (75.0 - 131.1) & 108.2 (82.3 - 153.5) & 152.5 (106.2 - 235.5) \\
China      & 1,330 & 26.3 (19.1 - 37.5)  & 26.9 (19.8 - 38.3)  & 30.2 (21.8 - 45.0)   & 42.9 (28.3 - 67.0)    \\
Bangladesh & 156   & 7.6 (5.5 - 10.6)    & 8.0 (5.7 - 11.1)    & 8.9 (6.5 - 14.0)     & 14.0 (8.9 - 23.5)     \\
India      & 1,173 & 6.8 (5.2 - 9.1)     & 7.1 (5.3 - 9.4)     & 8.0 (5.8 - 11.4)     & 11.5 (8.0 - 18.5)     \\
Indonesia  & 243   & 5.4 (3.8 - 8.0)     & 5.6 (4.0 - 8.4)     & 6.3 (4.6 - 9.9)      & 9.8 (6.1 - 16.6)      \\
Vietnam    & 90    & 11.5 (9.1 - 15.1)   & 11.7 (9.3 - 15.5)   & 12.9 (10.1 - 17.0)   & 17.0 (12.7 - 23.9)   \\ \\
\\
\multicolumn{6}{l}{{Current population occupying land exposed to inundation under 2300 RSL projections}}   \\\hline
Region         & Total Pop.            & RCP 2.6/K14        & RCP 2.6/DP16       & RCP 8.5/K14        & RCP 8.5/DP16        \\ \hline
World      & 6,836 & 165.9 (55.4 - 495.9) & 167.0 (82.8 - 357.9) & 306.4 (102.1 - 704.0) & 950.4 (765.6 - 1,162.6) \\
China      & 1,330 & 41.5 (11.2 - 125.6)  & 41.6 (17.3 - 97.8)   & 79.9 (20.7 - 171.1)   & 207.4 (171.1 - 251.9)   \\
Bangladesh & 156   & 29.7 (11.2 - 70.7)   & 29.9 (16.8 - 54.5)   & 49.8 (20.1 - 95.1)    & 117.0 (101.5 - 129.2)   \\
India      & 1,173 & 12.7 (4.3 - 42.8)    & 12.8 (6.5 - 28.3)    & 25.1 (8.0 - 71.1)     & 105.3 (78.1 - 132.5)    \\
Indonesia  & 243   & 9.1 (1.7 - 35.7)     & 9.0 (2.9 - 24.8)     & 19.6 (4.3 - 50.8)     & 65.7 (52.9 - 81.0)      \\
Vietnam    & 90    & 16.2 (5.4 - 38.7)    & 16.2 (8.0 - 31.4)    & 27.4 (9.6 - 49.5)     & 57.0 (50.7 - 61.4)     \\ \\ 
\hline
\multicolumn{6}{l}{Population currently living in land at risk of permanent inundation  based on median (5th--95th percentile) RSL projections.} \\
\multicolumn{6}{l}{Population densities based on 2010 estimates.} \\
\multicolumn{6}{l}{The top five countries with the most exposure in 2300 under median RCP 8.5/DP16 are included in this table.}
\end{tabular}
\label{tab:totalpopulationexposure} 

\end{table*}

\begin{figure}[tb]
\centering
\includegraphics[width=20pc]{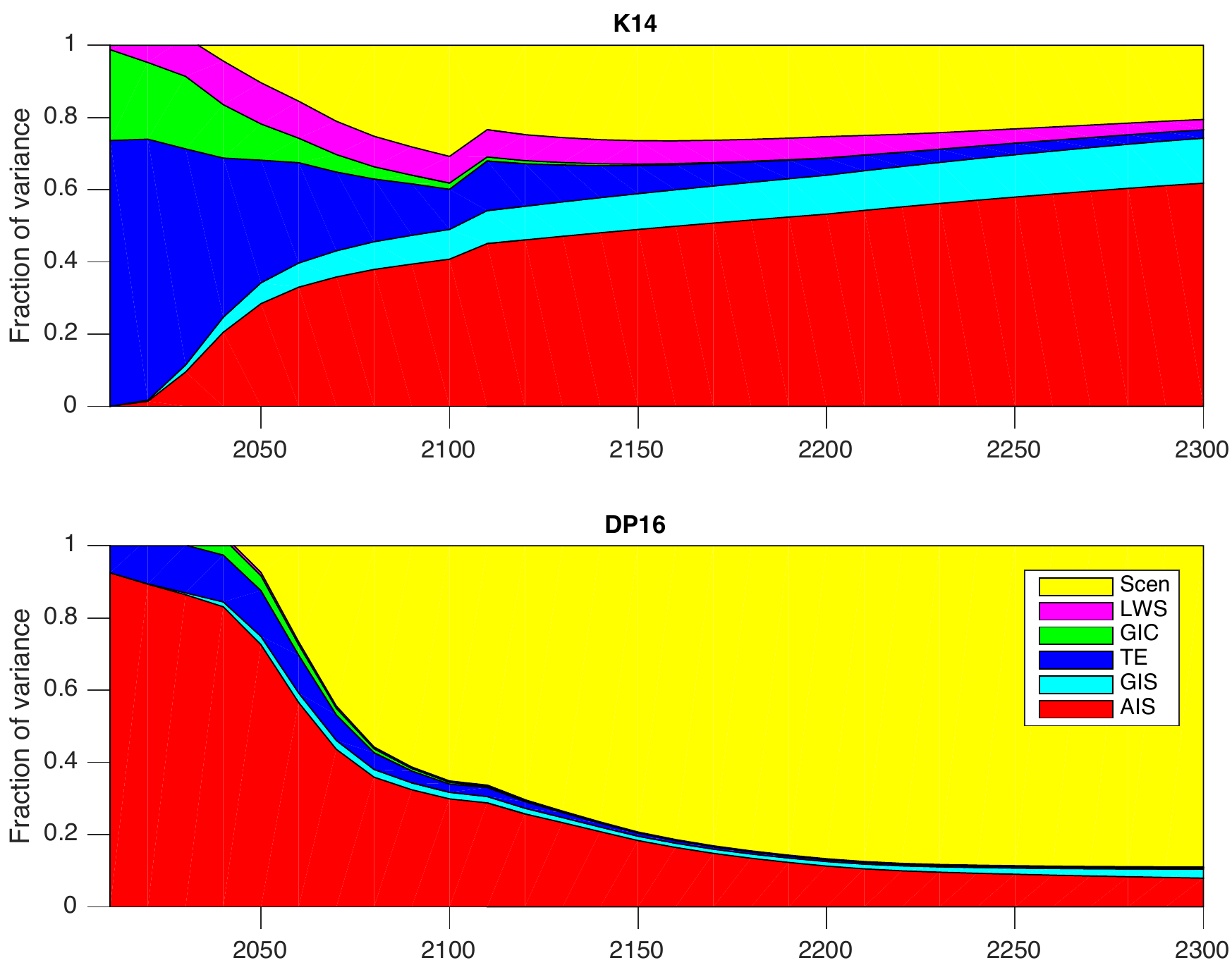}
\caption{Fractional contributions to the variance in GMSL projections over time under K14 (top) and DP16 (bottom) from:  Antarctic ice sheet (AIS; red),  Greenland ice sheet (GIS; cyan), thermal expansion (TE; blue), glaciers and ice cap (GIC; green), land water storage (LWS; magenta), and scenario uncertainty (Scen; yellow). Variances are calculated from bottom to top, so (for example) the top of the GIS wedge is the variance in the summed contributions of AIS and GIS to GMSL. All components are taken from RCP 4.5 except for the final total variance including scenario uncertainty, which is based on equally weighted pooled, projections for RCP 2.6, 4.5, and 8.5. The discontinuity between 2100 and 2110 is due to the reduction in the number of CMIP5 model simulations available beyond 2100.}
\label{fig:variance}
\end{figure}

As a consequence of this difference, the proportion of total projection variance attributable to emissions changes significantly with the incorporation of the DP16 ensemble (Figure \ref{fig:variance}). Under K14, relative to RCP 4.5, thermal expansion is initially the dominant contributor to projection variance (accounting for about 70\% of total variance in 2020). By 2060, AIS accounts for one-third of total variance and is the single largest contributor. The AIS share grows over time, accounting for more than 60\% of total variance by 2300. Assuming RCP 2.6, 4.5 and 8.5 are all treated as equally likely, scenario uncertainty accounts for only $\about 10\%$ of total variance in 2050, a share that grows to 20--30\% by 2070 and stays in this range through 2300.

Under DP16, physical uncertainty in AIS initially dominates total variance (89\% in 2020). This share declines over time, predominantly losing out to emissions scenario uncertainty, which grows from 8\% of total variance in 2050, to 45\% in 2070,  to 65\% in 2100, and continues to grow to 89\% in 2250.  This shift reflects the larger sensitivity of the DP16 AIS projections to emissions scenario.

\begin{figure}[tb]
\centering
\includegraphics[width=20pc]{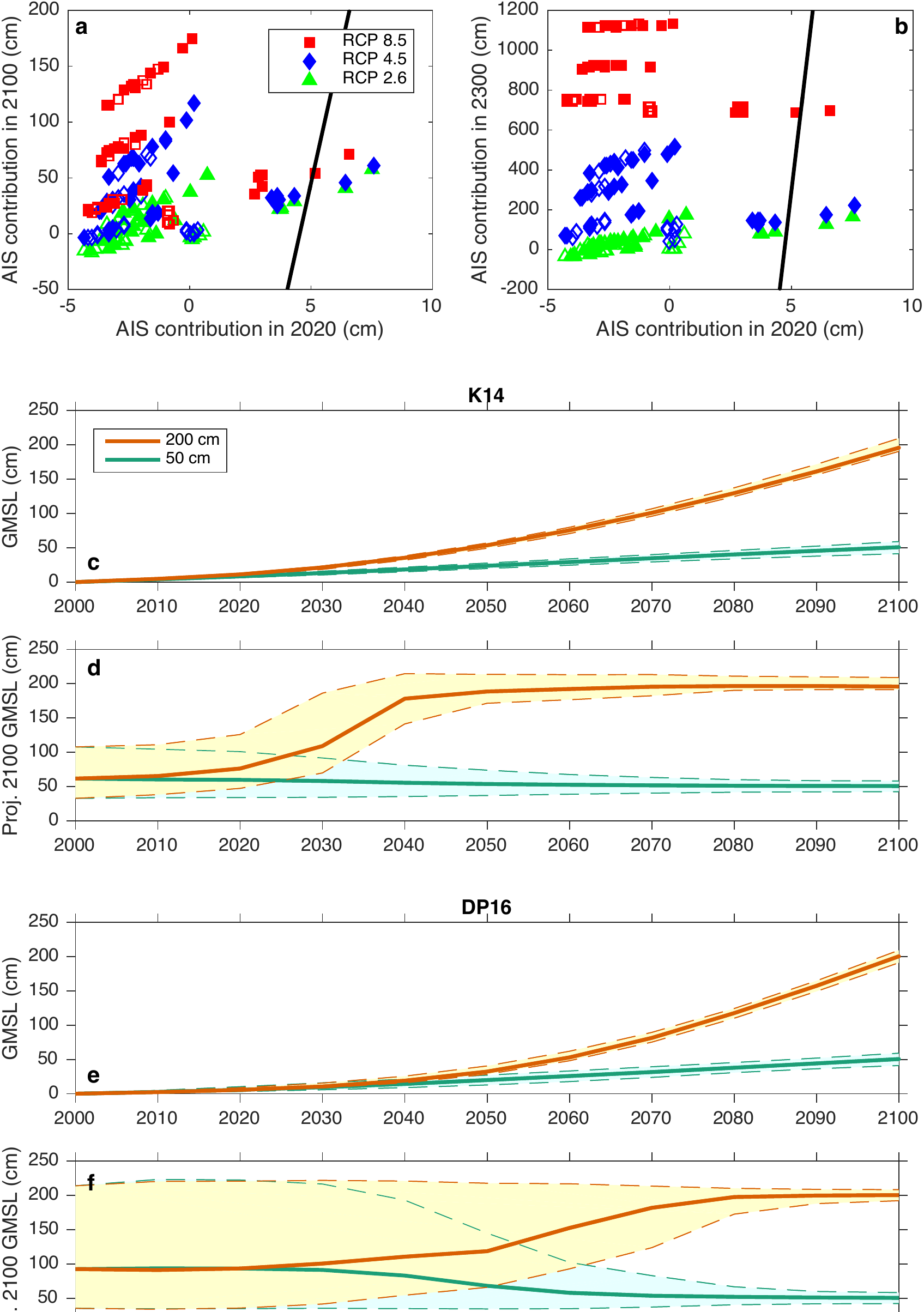}
\caption{(a-b) Relationship between the Antarctic ice sheet contribution to GMSL in 2020 and that in (a) 2100 or (b) 2300. Black line is the relationship in the K14 projections. Red/blue/green is the DP16 ensemble (red = RCP 8.5; blue = RCP 4.5; green = RCP 2.6; filled = with bias correction; open = without bias correction). (c, e) GMSL projections consistent with $50 \pm 10$ cm (green) and $200 \pm 10$ cm (orange) of GMSL rise in 2100 under (c) K14 and (e) DP16. (d, f) GMSL projections for 2100 conditional on observations in a given decade falling within  the bounds of the 50 cm (green) or 200 cm (orange) time paths. In (c-f),  heavy line = median; dashed/shaded region = 5th--95th percentile.}
\label{fig:AIScontrib_t2vst1}
\end{figure}

The assumption of a simple linear change in rate of mass loss underlying K14 leads to a perfect correlation between the rate of AIS mass loss observed in the near term and that projected for the long term (Figure \ref{fig:AIScontrib_t2vst1}a,b). If this assumption were correct, knowing that AIS mass loss in the first decades of this century fell in the middle of the estimated distribution would rule out high-end mass loss late in the century or beyond. By contrast,  DP16 projections reveal no correlation between the AIS contribution to GMSL in 2020 and that in 2100 ($r = -0.08$, pooling across RCPs and both with and without an ocean temperature adjustment), and only a weak correlation between the AIS contribution in 2050 and in 2100 ($r = 0.26$). In the second half of the century, observed AIS behavior becomes more strongly predictive of long-term behavior; the correlation with the AIS contribution to 2300 grows from $r = 0.26$ in 2050 to $r = 0.82$ in 2100, $r = 0.97$ in 2150, and $r = 0.997$ in 2200. The general lack of correlation between the AIS contributions in 2020 and 2100, and the rapidly strengthening correlations in the second half of the 21st century, are caused by a transition in the model from an ocean-dominated driver of ice-shelf loss (and reduced buttressing) to an atmosphere-dominated driver via hydrofracturing. 

The lack of correlation between  early 21st century and subsequent projections has important implications for the ability of GMSL observations to constrain future GMSL rise. In  K14, simulations consistent with $50 \pm 10$ cm of GMSL rise in 2100 diverge from those consistent with $200 \pm 10$ cm of GMSL rise by the 2020s (Figure \ref{fig:AIScontrib_t2vst1}c). The median conditional projections for 2100 under the 200 cm scenario exceed the 95th percentile under the 50 cm scenario by 2030, and the 5th percentile of the 200 cm conditional distribution exceeds the 95th percentile under the 50 cm scenario shortly thereafter. 95\% of projections for 2100 under the 200 cm time path exceed 100 cm in the 2030s and 150 cm in the 2040s (Figure \ref{fig:AIScontrib_t2vst1}d).  By contrast, the more complex temporal dynamics of the DP16 simulations  delays the divergence of the 50 and 200 cm time paths until around 2050 (Figure \ref{fig:AIScontrib_t2vst1}e). The median conditional projection for 2100 under the 200 cm scenario does not exceed the 95th percentile of the 50 cm scenario until the 2050s, with the 5th percentile of the 200 cm conditional distribution exceeding the 95th percentile under the 50 cm scenario in the 2060s.  95\% of projections for 2100 under the 200 cm time path exceed 50 cm in the 2040s,  100 cm in the 2060s, and  150 cm in the 2070s (Figure \ref{fig:AIScontrib_t2vst1}f).

\begin{figure}[tb]
\centering
\includegraphics[width=20pc]{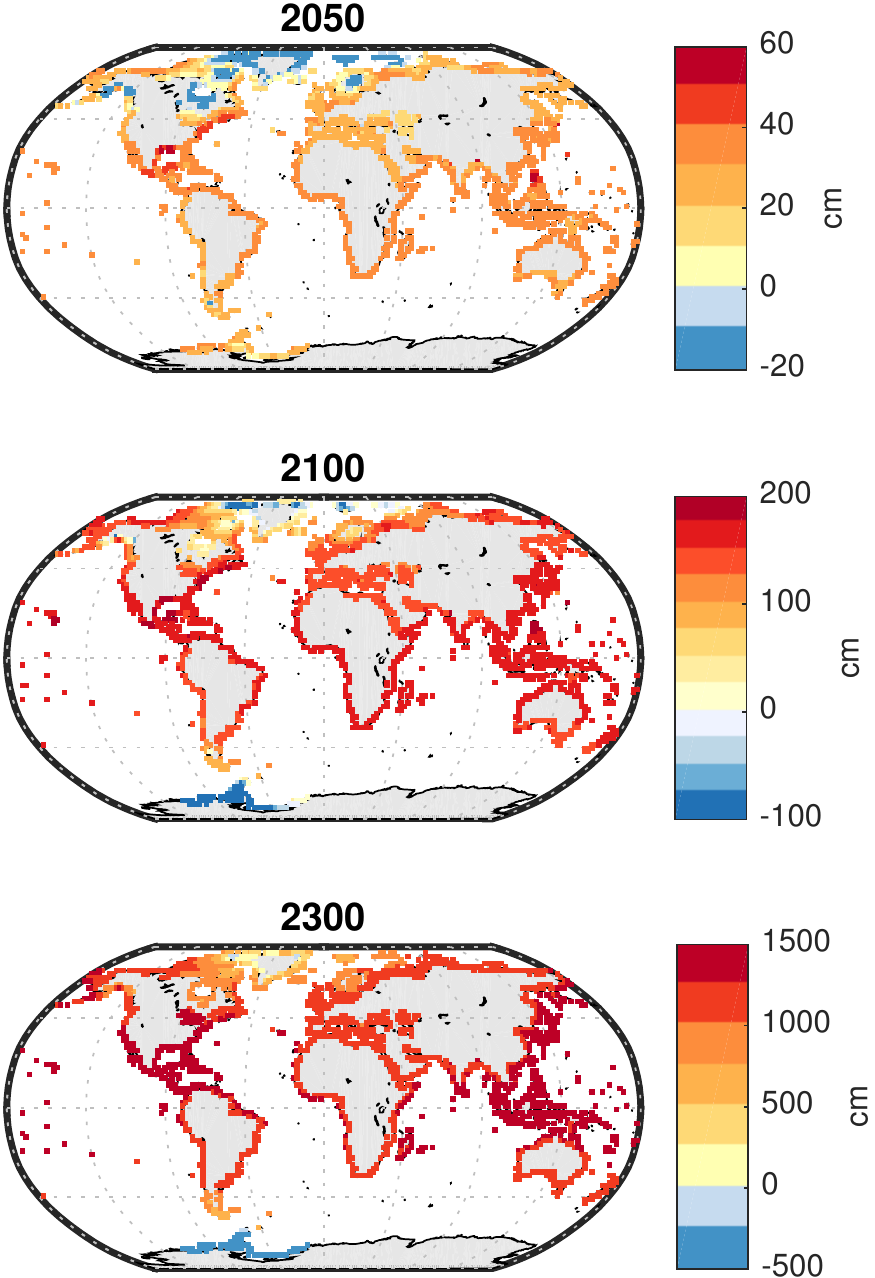}
\caption{Median DP16 RSL projections under RCP 8.5 in 2050, 2100 and 2300.}
\label{fig:LSLproj_Pl_5-15}
\end{figure}

The effect of DP16 on RSL projections is as would be expected based on the change in projected WAIS and EAIS contributions and their associated static-equilibrium fingerprints (Figure \ref{fig:LSLproj_Pl_5-15} and Supporting Information). Relative to K14, the effect on RSL projections for 2050 is minimal ($< 4$ cm).  By 2100, however, the increase in median ASL contribution decreases projected RSL rise in the Antarctic, while enhancing it most strongly in in a geographic swath including North America, the central Pacific, Australia, southeast Asia, and parts of India and Africa same. (Detailed simulation frequency distributions of RSL at tide gauge sites and on the global coastal grid are provided in the Supporting Information.) 

\section{Discussion and Conclusions}

The replacement of the probabilistic, expert-assessment- and expert-elicitation-based AIS projections of K14 with the physical-model-based projections of DP16 leads to a number of significant effects on GMSL and RSL projections. 

First, the use of explicit physics including novel ice-shelf-hydrofracturing and ice-cliff-collapse mechanisms in DP16 leads to a significant upward shift in central projections for strong (RCP 8.5) and moderate (RCP 4.5) forcing scenarios. The DP16 simulations provide physically meaningful pathways that can lead to $>2.0$ m of total GMSL rise by 2100 under RCP 8.5 and $>1.5$ m under RCP 4.5 \citep{Oppenheimer2016a,Sweet2017a}.

Second, in the second half of this century and beyond, sea-level projections incorporating the DP16 ensemble are significantly more forcing-sensitive than the K14 projections. Due primarily to the significant number of simulations involving collapse of parts of AIS under strong forcing, the gap in the median GMSL projection for 2100 between RCP 8.5 and RCP 2.6 grows from 30 cm under K14 to 90 cm under the projections under DP16. Under RCP 2.6 and DP16, the 99th percentile projection remains below 2 m through 2200. If these findings are correct, they point to a significantly larger mitigation benefit than indicated by the AR5 or K14 sea-level projections.

Third, the DP16 projections indicate a much weaker correlation between the near-term behavior of AIS and its contribution to GMSL rise over the course of this century and beyond. Finding the planet on a `moderate' sea-level rise pathway over the first half of the 21st century thus cannot exclude `extreme' outcomes subsequently. For end-users employing discrete scenarios of sea-level rise, such as those constructed by \citet{Sweet2017a}, this means that `extreme' future scenarios need to be considered even if they overestimate current rates of sea-level rise. Constraining the future behavior of the AIS requires more detailed process-based modeling than the simple relationships used by \citet{Bamber2013a}, K14, and others would indicate. 

That said, end-users of sea-level rise projections should be cognizant of single-study bias: the DP16 simulations should be viewed as expanding scientific understanding of the space of the physically coherent, rather than as offering firm projections of what will be.  More robust projections of future Antarctic contributions to sea-level rise require a more through exploration of appropriate values for parameters such as the maximum rate of ice cliff retreat and the sensitivity of ice-shelves to hydrofracturing, as well as of uncertainty in the regional climate response to forcing. They also require more physically based representations of key processes, including  structural cliff failure. Currently, the potential for model intercomparisons is hampered by the lack of representations of these processes in most continental-scale ice-sheet models. 

Moreover, there remain important physical processes that are not currently in any continental-scale model but could be critical for the timing and pace of major ice-sheet retreat. For example, at present, continental-scale ice-sheet models poorly represent the meltwater-buffering capacity of firn, the transitional layer between newer snow and underlying ice. In the future, as summer air temperatures begin to drive the production of more rain and surface melt, meltwater will be absorbed by the firn layer as long as the layer contains uncompacted pore space, limiting the meltwater's potential to flow into crevasses and hydrofracture the underlying ice \citep{Munneke2014a}. 

The breadth of published projections, as well as of remaining structural uncertainties, highlight the fact that future sea-level rise remains an arena of deep uncertainty \citep{Kasperson2008a,Heal2014a}. For the foreseeable future, there will not be a single, uniquely valid approach for estimating the probability of different levels of future change. End-users should therefore consider applying robust decision frameworks and/or adaptive management frameworks  appropriate for deeply uncertain contexts. Where feasible, they could employ decision criteria that seek to minimize regret or optimize weighted mixtures of expected utilities across multiple possible  distributions  \citep[e.g.,][]{Heal2014a}, rather than relying on a single distribution. 

They could also try to structure decisions in a staged fashion, such that the decisions being made today depend primarily upon mid-century projections while leaving open a variety of options for later in the century \citep[e.g.,][]{Ranger2013a}. The value of this approach stems from the  robustness of mid-century sea-level rise projections relative to those for later in the century. Through 2050, the K14 and DP16 projections overlap substantially, as do the projections within both sets for different RCPs (Table \ref{tab:GMSL}). 

One simple  multiple-probability-distribution approach involves giving special consideration to physically plausible but low-likelihood projections in the high-end tail of projected probability distributions \citep[e.g.,][]{Buchanan2016a}. K14 suggested that the 99.9th percentile of their projection for RCP 8.5, which yielded  $\about 2.5$ m of GMSL rise by 2100, represented a physically plausible `worst case.' The highest values from the DP16 projections are only modestly higher, although high values occur with greater frequency: under RCP 8.5, the DP16 ensemble revises the frequency of a $>2.5$ m GMSL rise upward from $0.1\%$ to $3\%$. However, even higher frequencies for $>2.5$ m outcomes are conceivable; the DP16 ensemble may not cover the full space of plausible outcomes.  Notably, rates of ice-cliff collapse faster than the 5 km/yr maximum of DP16 have been observed in parts of Greenland, and faster rates of ice-cliff collapse would yield faster rates of ice-sheet retreat.

Moving forward, the development of probabilistic ice-sheet models that incorporate ice-sheet instabilities in a manner consistent with more detailed ice-sheet models is a key challenge for sea-level risk assessment \citep[e.g.,][]{Wong2017a,Nauels2017a}.  Such probabilistic  models will enable estimation of sea-level rise probabilities that reflect emissions sensitivity and the potential for rapid increases in discharge rate more accurately than current approaches.  Probabilistic projections are a valuable input into the design of projects and policies intended to manage coastal flood risk \citep[e.g.,][]{Buchanan2016a,Lempert2012a,Wong2017a,Oppenheimer2016a}, as well as assessments of the value of climate change mitigation \citep[e.g.,][]{Hinkel2014a,Houser2015a,Diaz2016a}. They will also enable value-of-information analyses, which can inform the design of observation systems intended to reduce the key physical uncertainties underlying  future sea-level projections \citep{Cooke2014a}. 

Probabilistic assessment also requires more research on potential bounds for factors that influence the AIS. For example, it is unclear whether even a full ensemble of GCMs would fully constrain the range of plausible distributions of near surface ocean temperature, sea ice, and storm tracks near the AIS.  More research is also needed on interactions and feedbacks across the AIS and between the AIS and the rest of the world, including through  poorly understood mechanisms like ocean circulation that could over long time scales influence both ice-sheet retreat and other drivers of coastal flood risk (such as tropical cyclones) around the globe. 


%

\acknowledgments
REK was supported in part by a grant from Rhodium Group (for whom he has previously worked as a consultant), as part of the Climate Impact Lab consortium, and in part by a grant from the National Science Foundation (ICER-1663807). RMD and DP were supported by grants from the National Science Foundation (OCE-1202632 PlioMAX, AGS 1203910/1203792, and ICER-1664013/1663693).
This paper is a contribution to PALSEA2 (Palaeo-Constraints on Sea-Level Rise), which is a working group of Past Global Changes/IMAGES (International Marine Past Global Change Study) and an International Focus Group of the International Union for Quaternary Research. We thank Climate Ready Boston for inspiring this paper.
 We acknowledge the World Climate Research Programme's Working Group on Coupled Modeling, which is responsible for CMIP, and we thank the climate modeling groups (listed in Supporting Information) for producing and making available their model output. For CMIP, the U.S. Department of Energy's Program for Climate Model Diagnosis and Intercomparison provides coordinating support and led development of software infrastructure in partnership with the Global Organization for Earth System Science Portals. 
 Code for generating sea-level projections is available in the ProjectSL (https://github.com/bobkopp/ProjectSL) and LocalizeSL (https://github.com/bobkopp/LocalizeSL) repositories on Github.




\newpage
\clearpage

\setcounter{table}{0}
\setcounter{figure}{0}
\setcounter{page}{1}

\renewcommand{\thepage}{S\arabic{page}}
\renewcommand{\thetable}{S\arabic{table}}
\renewcommand{\thefigure}{S\arabic{figure}}

\supportinginfo{Evolving understanding of Antarctic ice-sheet physics and ambiguity in probabilistic sea-level projections}

\authors{Robert E. Kopp\affil{1}, Robert M. DeConto\affil{2}, Daniel A. Bader\affil{3}, Carling C. Hay\affil{1,4,8}, Radley M. Horton\affil{3}, Scott Kulp\affil{5}, Michael Oppenheimer\affil{6}, David Pollard\affil{7}, Benjamin H. Strauss\affil{5}}

\affiliation{1}{Department of Earth \& Planetary Sciences and Institute of Earth, Ocean \& Atmospheric Sciences, Rutgers University, New Brunswick, NJ, USA}
\affiliation{2}{Department of Geosciences, University of Massachusetts--Amherst, Amherst, MA, USA}
\affiliation{3}{Center for Climate Systems Research, Columbia University, New York, NY, USA}
\affiliation{4}{Department of Earth \& Planetary Sciences, Harvard University, Cambridge, MA, USA}
\affiliation{5}{Climate Central, Princeton, NJ, USA}
\affiliation{6}{Woodrow Wilson School of Public \& International Affairs and Department of Geosciences, Princeton University, Princeton, NJ, USA}
\affiliation{7}{Earth and Environmental Systems Institute, Pennsylvania State University, University Park, PA, USA}
\affiliation{8}{\emph{Current address:} Boston College, Boston, MA, USA}

\correspondingauthor{R. E. Kopp}{robert.kopp@rutgers.edu}

\section*{Contents}
\begin{enumerate}
\item Text S1
\item Data Set S1
\item Data Set S2
\item Data Set S3
\item Data Set S4
\item Data Set S5
\item Data Set S6
\end{enumerate}

\section*{Text S1: Population Exposure Assessment Detailed Methods.}

To assess topography as required for this analysis, we employ the NASA SRTM digital elevation model, which is based on a radar satellite mission in 2000. This Data Set has nearly global coverage (including latitudes 60N--54S, covering land inhabited by more than 99.9 percent of global population), and is available at a 1 arcsec (SRTM 3.0) horizontal resolution \citep{NASA-JPL2013a}. SRTM, as distributed, is referenced to the EGM96 geoid (henceforth denoted by $SRTM_{EGM96}$) at a 1 m vertical resolution, with RMSE less than 10 m \citep{LaLonde2010a,Rodriguez2005a}. Derived from radar measurements, $SRTM_{EGM96}$ is an unclassified (surface) elevation model, and significant positive bias is expected in regions of dense urban development and vegetation \citep{Shortridge2011a}.

To convert elevations to a tidal vertical datum, we use the global mean sea level (MSL) model MSS\_CNES\_CLS\_11 \citep{AVISO2011a}, based on TOPEX/Poseidon satellite measurements, and referenced to the GLAS ellipsoid ($MSL_{GLAS}$). We also employ mean higher-high water ($MHHW_{MSL}$) deviations from MSL provided by Mark Merrifield, University of Hawaii, developed using the model TPX08 \citep{Egbert2002a}. We convert MSL and SRTM to a common ellipsoidal datum (WGS84) using NOAA's VDatum tool \citep{Parker2003a} version 3.7, and subtract the MHHW grids from $MSL_{WGS84}$ to find $MHHW_{WGS84}$. We then subtract this tidal grid from $SRTM_{WGS84}$ to produce our final elevation map, $SRTM_{MHHW}$.

We resample a given relative sea-level rise projection grid $X$ using bilinear interpolation to match the horizontal resolution of SRTM, and threshold elevation against these water heights to produce the inundation surface $THRESH_x$. Hydrological connectivity to the ocean is typically enforced in such analyses, but we find that high-frequency errors present in SRTM create significant speckle noise in the flood maps, causing some truly connected areas to appear isolated. We instead perform connected components analysis at the 20m water height above MHHW, producing surface $CONTIG_{20m}$, and perform the intersection $HYBRID_x=THRESH_x\cap CONTIG_{20m}$. Isolated, low-lying land separated from ocean by at least a 20m high ridge is therefore removed from the final surface, $HYBRID_x$, while low-lying ocean-side land is included, reducing sensitivity to speckle noise. We compute inundation surfaces given different projection models (K14 and DP16), emissions scenarios (RCP 2.6 and RCP 8.5), percentiles of RSL projections (5th, 50th, and 95th), and years (2100 and 2300).

To assess population exposure, we employ the LandScan 2010 High Resolution global Population Data Set, which provides total estimated populations living in 1 km square cells \citep{Bright2011a}. We refine this data using the SRTM Water Body Data Set (SWBD), which defines land cells at 1-arcsecond resolution. We resample Landscan at 1-arcsecond resolutions to align with the SRTM grids, assuming zero population in water cells, while proportionally increasing the population density in land cells to ensure total population in each 1 km square remains unchanged. We integrate exposure under each inundation surface and tabulate according to national boundaries defined by the Global Administrative Areas (GADM) 2.0 Data Set \citep{GADM2012a}.

Linked to the positive bias in SRTM elevation data, a notable negative bias has been detected in coastal population flood exposure analysis based on SRTM, at least within the United States, where higher quality, LIDAR-based elevation models are freely available for comparison \citep{Kulp2016a}.  LIDAR-based US national exposure estimates are $\about 45\%$ higher than SRTM-based estimates at 1 m above MHHW; $\about 150\%$ higher at 2~m and 3~m; and monotonically decline  to $\about 33\%$ higher at 10~m \citep{Kulp2016a}. The exposure values presented here can thus be viewed as likely to significantly underestimate the true hazard. We nevertheless include them to provide some illustration of the ramifications of different projections. SRTM data are widely used in analysis of global  exposure to sea level rise and coastal flooding \citep[e.g.,][]{Hinkel2014a}, and the major available alternative, the Global Land One-kilometer Base Elevation (GLOBE) gridded elevation model, is far coarser in resolution and based on underlying data of varying and unknown quality by region \citep{Kulp2016a}. 

The LandScan 2010 High Resolution global Population Data Set  is copyrighted by UT-Battelle, LLC, operator of Oak Ridge National Laboratory under Contract No. DE-AC05-00OR22725 with the United States Department of Energy.  The United States Government has certain rights in this Data Set.  Neither UT-Battelle, LLC nor the United States Department of Energy, nor any of their employees, makes any warranty, express or implied, or assumes any legal liability or responsibility for the accuracy, completeness, or usefulness of the data set.

\section*{Data Set S1}

Data Set S1 provides time series of WAIS, EAIS and total AIS contributions to GMSL from 2000 to 2300 from DP16.

\section*{Data Set S2}

Data Set S2 provides estimated non-climatic background rates and IDs, latitude, and longitude of tide-gauge and grid point locations.

\section*{Data Set S3}
Data Set S3 provides K14 GMSL and RSL projections at decadal intervals from 2010 to 2300.

\section*{Data Set S4}

Data Set S4 provides DP16 GMSL and RSL projections at decadal intervals from 2010 to 2300.

\section*{Data Set S5}
Data Set S5 provides DP16 GMSL and RSL projections at decadal intervals from 2010 to 2300 without bias correction.

\section*{Data Set S6}
Data Set S6 provides current population by country occupying land exposed to inundation under K14 and DP16 projected 2100 and 2300 RSL change.


\begin{table*}[h]
\caption{CMIP5 models used for thermal expansion and oceanographic processes}
\centering
\scriptsize{
    \begin{tabular}{r|rrr|rrr}
Model & \multicolumn{3}{c}{Oceanographic} & \multicolumn{3}{c}{GIC} \\
 & RCP 8.5 & RCP 4.5 & RCP 2.6 & RCP 8.5 & RCP 4.5 & RCP 2.6 \\ \hline
access1-0 & 21 & 21 &&& \\
access1-3 & 21 & 21 &&&  \\
bcc-csm1-1 & 23 & 23 & 23 & 23 & 23 & 23 \\
bcc-csm1-1-m & 21 & 21 & 21 \\
canesm2 & 21 & 23 & 23 & 21 & 23 & 23 \\
ccsm4 & 21 & 21 & 21 & 21 & 21 & 21 \\
cmcc-cesm & 21 && &&& \\
cmcc-cm & 21 & 21 &&&  \\
cmcc-cms & 21 & 21 &&& \\
cnrm-cm5 & 23 & 23 & 21 & 23 & 23 & 21 \\
csiro-mk3-6-0 & 21 & 21 & 21 & 23 & 23 & 21 \\
gfdl-cm3 & 21 & 23 & 21 & 21 & 21 & 21 \\
gfdl-esm2g & 21 & 21 & 21 \\
gfdl-esm2m & 21 & 21 & 21 \\
giss-e2-r & 23 & 23 & 23 & 23 & 23 \\
giss-e2-r-cc & 21 & 21 &&& \\
hadgem2-cc & 21 & &&& \\
hadgem2-es & 21 &  & 23 & 23 & 23 & 23 \\
inmcm4 & 21 & 21 &  & 21 & 21 \\
ipsl-cm5a-lr & 23 & 23 & 23 & 23 & 23 & 23 \\
ipsl-cm5a-mr & 21 & 23 & 21 \\
miroc-esm & 21 & 23 & 21 & 21 & 21 & 21 \\
miroc-esm-chem & 21 & 21 & 21 \\
miroc5 &  &  &  & 21 & 21 & 21 \\
mpi-esm-lr & 23 & 23 & 23 & 23 & 23 & 23 \\
mpi-esm-mr & 21 & 21 & 21 \\
mri-cgcm3 & 21 &  & 21 & 21 & 21 & 21 \\
noresm1-m & 21 & 23 & 21 & 21 & 23 & 21 \\
noresm1-me & 21 & 21 & 21 \\
    \hline
    \multicolumn{7}{l}{\scriptsize{21 = to 2100, 23 = to 2300.}}
    \end{tabular}}
    \label{Stab:CMIP5}
\end{table*}

\begin{table*}
\caption{Projections of GMSL rise without bias correction (cm)}
\centering
{\scriptsize
\begin{tabular}{l l l l l l }
 & 50 & 17--83 & 5--95 & 1--99 & 99.9 \\
\hline
RCP 8.5 \\
2050 & 26 & 20--33 & 16--39 & 12--45 & 51 \\
2100 & 134 & 89--202 & 76--224 & 64--244 & 277 \\
2200 & 713 & 579--892 & 530--959 & 495--1047 & 1193 \\
2300 & 1169 & 981--1410 & 913--1553 & 861--1752 & 2007 \\
RCP 4.5 \\
2050 & 22 & 16--28 & 12--33 & 9--38 & 42 \\
2100 & 74 & 51--115 & 41--134 & 33--149 & 168 \\
2200 & 241 & 134--386 & 104--439 & 79--489 & 571 \\
2300 & 397 & 224--582 & 164--680 & 118--789 & 983 \\
RCP 2.6 \\
2050 & 20 & 15--25 & 11--29 & 8--33 & 37 \\
2100 & 44 & 31--62 & 23--75 & 15--87 & 101 \\
2200 & 86 & 58--123 & 40--153 & 25--194 & 267 \\
2300 & 109 & 66--172 & 39--238 & 16--336 & 492 \\
\hline
\end{tabular}}
\label{Stab:GMSL_noBC}
\end{table*}

\begin{table}[tb]
\caption{Projections of GMSL rise under different  assumptions regarding ice-cliff collapse (cm)}
\centering
\scriptsize
\begin{tabular}{l l l l l l}
  & 50 & 17--83 & 5--95 & 1--99 & 99.9 \\
\hline
\multicolumn5l{No Ice Cliff Collapse (VCLIF = 0 km/yr)} \\
RCP 8.5 \\
2050 & 35 & 31--39 & 28--43 & 25--47 & 54 \\
2100 & 124 & 109--139 & 97--154 & 87--172 & 213 \\
2200 & 603 & 560--661 & 532--722 & 507--810 & 986 \\
2300 & 994 & 916--1119 & 874--1262 & 840--1474 & 1622 \\
RCP 4.5 \\
2050 & 33 & 29--37 & 26--40 & 24--43 & 46 \\
2100 & 84 & 74--96 & 66--106 & 58--116 & 137 \\
2200 & 196 & 164--235 & 142--273 & 122--320 & 420 \\
2300 & 295 & 232--374 & 194--454 & 161--569 & 800 \\
RCP 2.6 \\
2050 & 32 & 29--35 & 26--38 & 24--41 & 43 \\
2100 & 66 & 58--75 & 52--83 & 46--92 & 108 \\
2200 & 129 & 111--156 & 101--186 & 93--225 & 301 \\
2300 & 170 & 138--227 & 123--292 & 111--388 & 554 \\
\hline
\multicolumn5l{Slow Ice Cliff Collapse (VCLIF = 1 km/yr)} \\
RCP 8.5 \\
2050 & 23 & 18--30 & 14--37 & 11--43 & 48 \\
2100 & 109 & 92--126 & 82--140 & 72--159 & 199 \\
2200 & 621 & 580--679 & 551--742 & 521--831 & 988 \\
2300 & 1032 & 953--1157 & 909--1302 & 865--1514 & 1655 \\
RCP 4.5 \\
2050 & 19 & 14--27 & 11--36 & 7--41 & 46 \\
2100 & 64 & 49--82 & 40--94 & 31--108 & 123 \\
2200 & 180 & 133--228 & 105--266 & 81--315 & 405 \\
2300 & 297 & 222--384 & 174--468 & 130--582 & 791 \\
RCP 2.6 \\
2050 & 17 & 13--24 & 10--34 & 6--38 & 43 \\
2100 & 38 & 27--55 & 19--75 & 12--87 & 100 \\
2200 & 72 & 46--117 & 32--151 & 21--191 & 263 \\
2300 & 96 & 50--173 & 27--240 & 9--332 & 534 \\
\hline
\multicolumn5l{Fast Ice Cliff Collapse (VCLIF = 5 km/yr)} \\
RCP 8.5 \\
2050 & 38 & 31--46 & 26--52 & 23--57 & 62 \\
2100 & 213 & 185--242 & 144--259 & 127--277 & 307 \\
2200 & 898 & 847--957 & 605--1021 & 559--1111 & 1265 \\
2300 & 1398 & 1303--1525 & 972--1668 & 882--1880 & 2031 \\
RCP 4.5 \\
2050 & 31 & 25--39 & 20--50 & 17--55 & 59 \\
2100 & 127 & 109--156 & 98--174 & 89--187 & 206 \\
2200 & 400 & 342--451 & 246--491 & 207--545 & 627 \\
2300 & 592 & 495--684 & 360--767 & 284--888 & 1070 \\
RCP 2.6 \\
2050 & 28 & 22--36 & 17--48 & 14--53 & 57 \\
2100 & 71 & 56--94 & 46--106 & 39--116 & 127 \\
2200 & 135 & 100--196 & 84--226 & 70--266 & 324 \\
2300 & 174 & 111--269 & 87--333 & 67--427 & 559 \\
\hline
\multicolumn5l{Columns indicate percentiles of simulation frequency distributions.}
\end{tabular}
\label{Stab:GMSLicesub-VCLIF}
\end{table}

\begin{table}[tb]
\caption{Projections of GMSL rise under different  assumptions regarding ice-shelf hydrofracturing (cm)}
\centering
\scriptsize
\begin{tabular}{l l l l l l}
  & 50 & 17--83 & 5--95 & 1--99 & 99.9 \\
\hline
\multicolumn5l{No Hydrofracturing (CREVLIQ = 0 m per (m/yr)$^{-2}$)} \\
RCP 8.5 \\
2050 & 39 & 33--49 & 29--54 & 26--59 & 66 \\
2100 & 125 & 107--147 & 96--163 & 86--183 & 226 \\
2200 & 588 & 545--646 & 520--709 & 499--802 & 985 \\
2300 & 980 & 902--1105 & 860--1248 & 826--1458 & 1612 \\
RCP 4.5 \\
2050 & 38 & 32--48 & 29--53 & 25--57 & 61 \\
2100 & 94 & 78--114 & 68--126 & 59--139 & 159 \\
2200 & 216 & 174--263 & 148--304 & 125--354 & 462 \\
2300 & 320 & 248--407 & 204--489 & 167--610 & 830 \\
RCP 2.6 \\
2050 & 36 & 31--47 & 28--51 & 25--55 & 58 \\
2100 & 79 & 65--99 & 57--110 & 49--120 & 135 \\
2200 & 157 & 123--193 & 108--224 & 97--267 & 344 \\
2300 & 213 & 161--275 & 136--340 & 118--441 & 613 \\
\hline
\multicolumn5l{Strong Hydrofracturing (CREVLIQ = 150 m per (m/yr)$^{-2}$)}\\
RCP 8.5 \\
2050 & 29 & 21--37 & 16--42 & 12--46 & 53 \\
2100 & 142 & 103--209 & 87--231 & 75--249 & 280 \\
2200 & 715 & 605--892 & 571--959 & 548--1047 & 1189 \\
2300 & 1166 & 995--1404 & 931--1551 & 886--1734 & 1982 \\
RCP 4.5 \\
2050 & 24 & 17--32 & 12--37 & 8--40 & 44 \\
2100 & 86 & 58--124 & 45--142 & 33--155 & 170 \\
2200 & 260 & 160--395 & 118--448 & 85--495 & 584 \\
2300 & 414 & 259--595 & 191--689 & 135--798 & 1002 \\
RCP 2.6 \\
2050 & 22 & 15--30 & 11--34 & 7--38 & 42 \\
2100 & 52 & 33--71 & 23--80 & 14--90 & 105 \\
2200 & 100 & 61--136 & 38--166 & 23--205 & 287 \\
2300 & 124 & 72--186 & 36--251 & 11--346 & 529 \\
\hline
\multicolumn5l{Columns indicate percentiles of simulation frequency distributions.}
\end{tabular}
\label{Stab:GMSLicesub-CREVLIQ}
\end{table}

\newpage
\clearpage

\begin{figure}[ht]
\centering
\includegraphics[width=20pc]{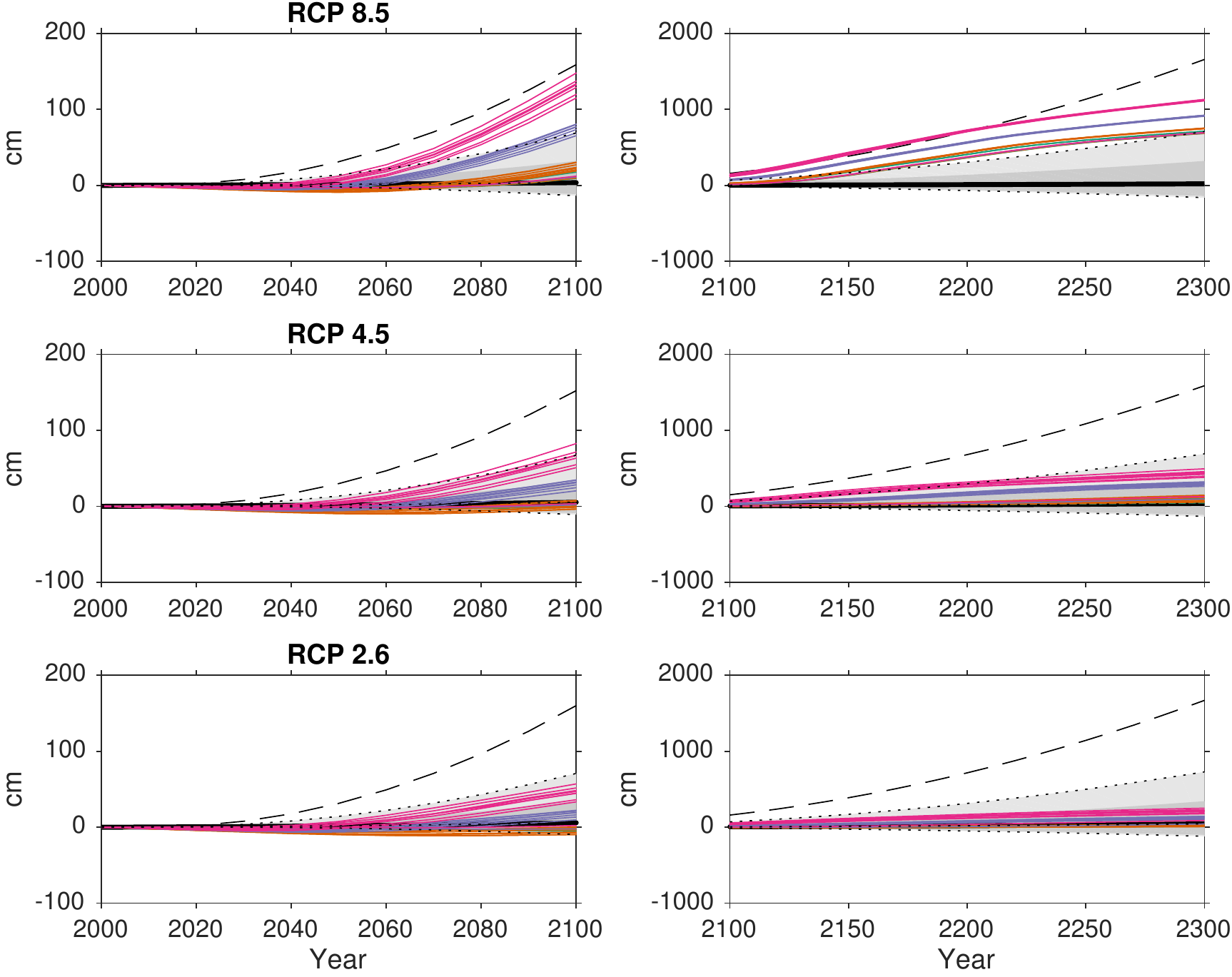}
\caption{Projections of the Antarctic ice sheet contribution to GMSL under the three RCPs without bias correction. As in Figure 2. }
\label{Sfig:AISoldandnew_noBC}
\end{figure}

\begin{figure}[ht]
\centering
\includegraphics[width=20pc]{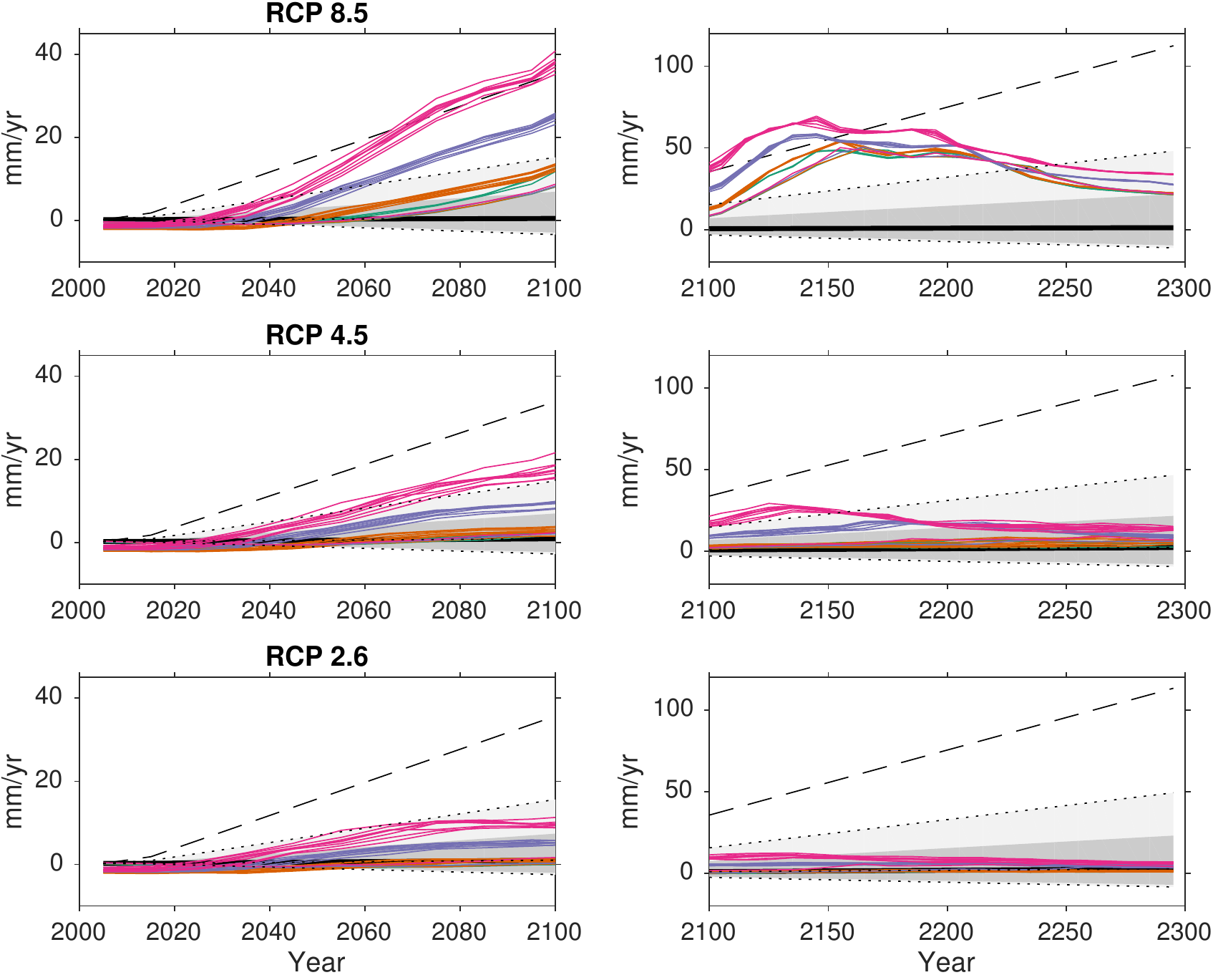}
\caption{Rates of contribution of the Antarctic ice sheet to GMSL under the three RCPs without bias correction. As in Figure 1.}
\label{Sfig:rate-AISoldandnew_noBC}
\end{figure}

\begin{figure}[ht]
\centering
\includegraphics[width=20pc]{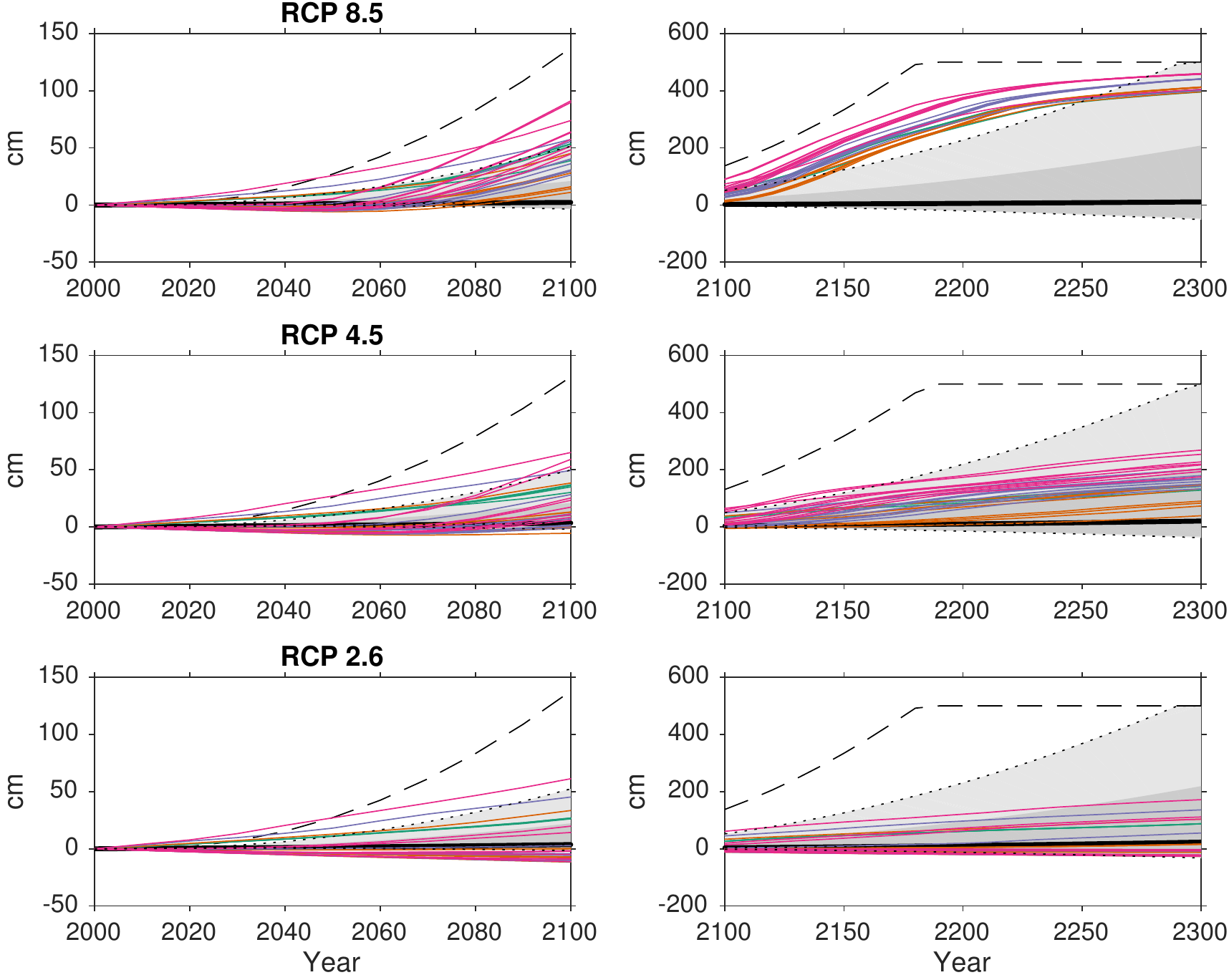}
\caption{West Antarctic ice sheet projections under the three RCPs. As in Figure 2.}
\label{Sfig:WAISoldandnew}
\end{figure}

\begin{figure}[ht]
\centering
\includegraphics[width=20pc]{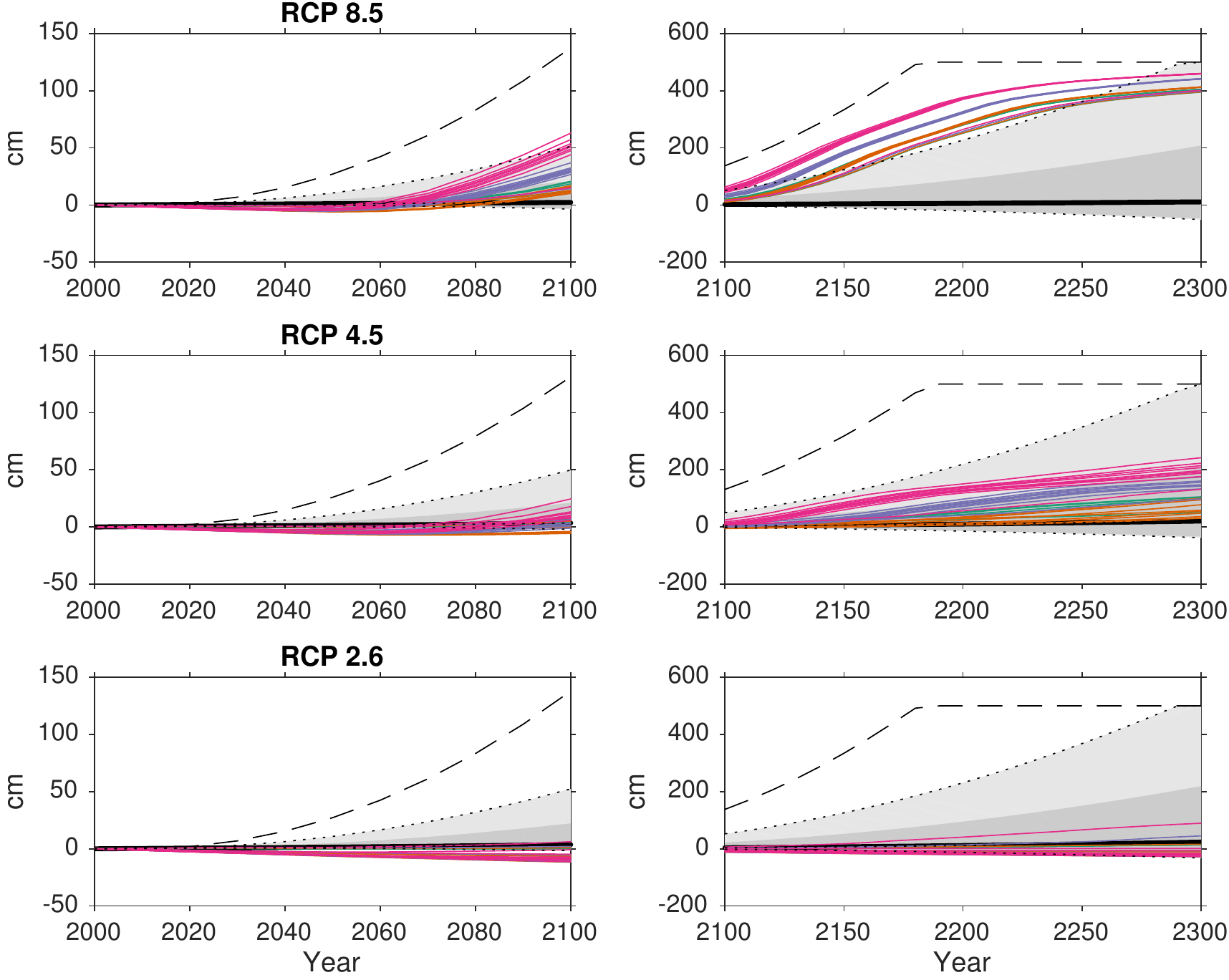}
\caption{West Antarctic ice sheet projections under the three RCPs without bias correction. As in Figure 2.}
\label{Sfig:WAISoldandnew_noBC}
\end{figure}

\begin{figure}[ht]
\centering
\includegraphics[width=20pc]{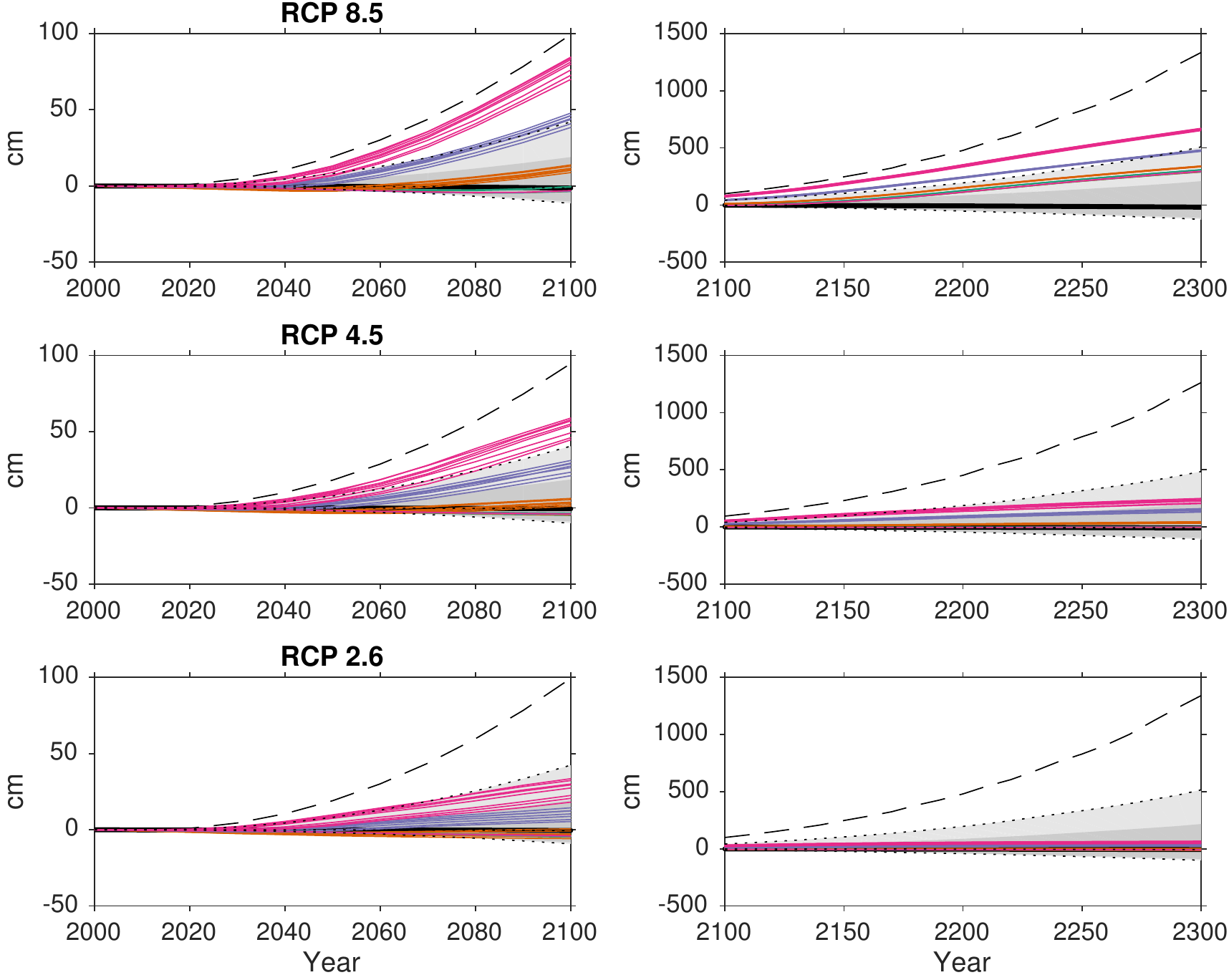}
\caption{East Antarctic ice sheet projections under the three RCPs. As in Figure 2.}
\label{Sfig:EAISoldandnew}
\end{figure}

\begin{figure}[ht]
\centering
\includegraphics[width=20pc]{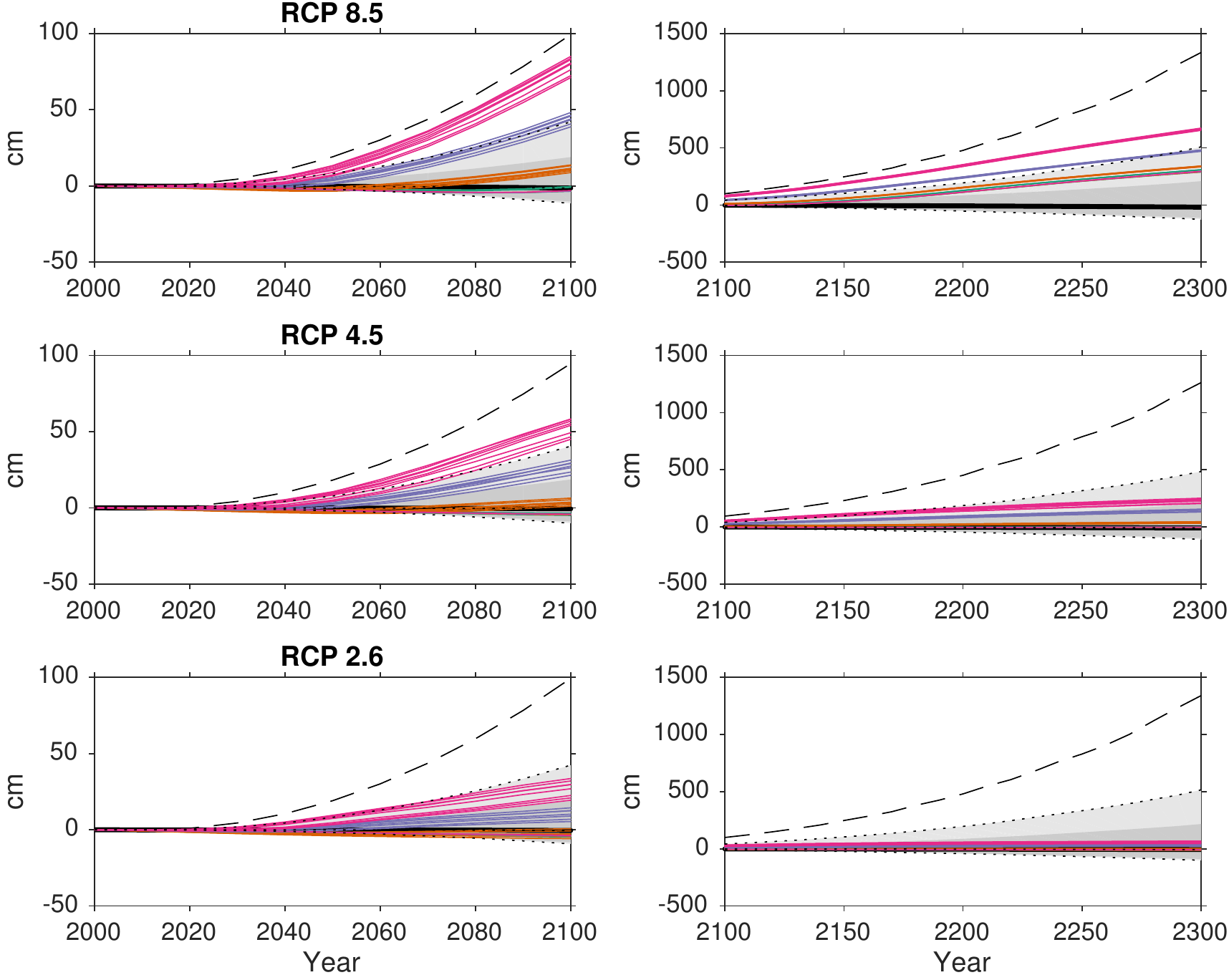}
\caption{East Antarctic ice sheet projections under the three RCPs without bias correction. As in Figure 2.}
\label{Sfig:EAISoldandnew_noBC}
\end{figure}

\begin{figure}[ht]
\centering
\includegraphics[width=20pc]{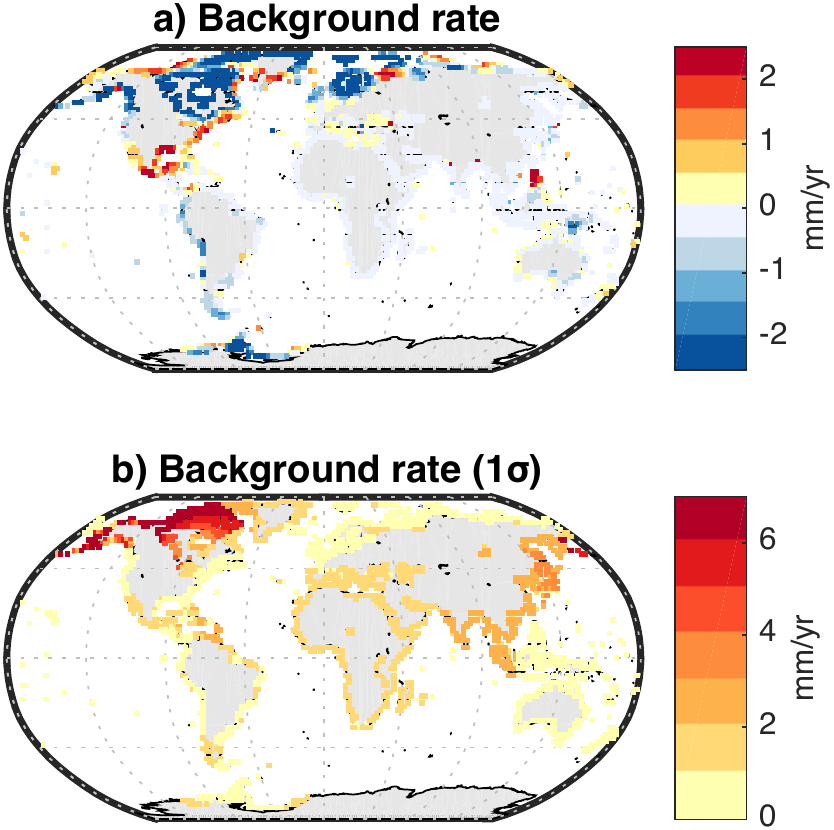}
\caption{Non-climatic background rates estimated on a grid using the spatio-temporal statistical model of tide-gauge data. (a) Mean estimate, (b) standard deviation.}
\label{Sfig:bkgdrate}
\end{figure}

\begin{figure}[ht]
\centering
\includegraphics[width=20pc]{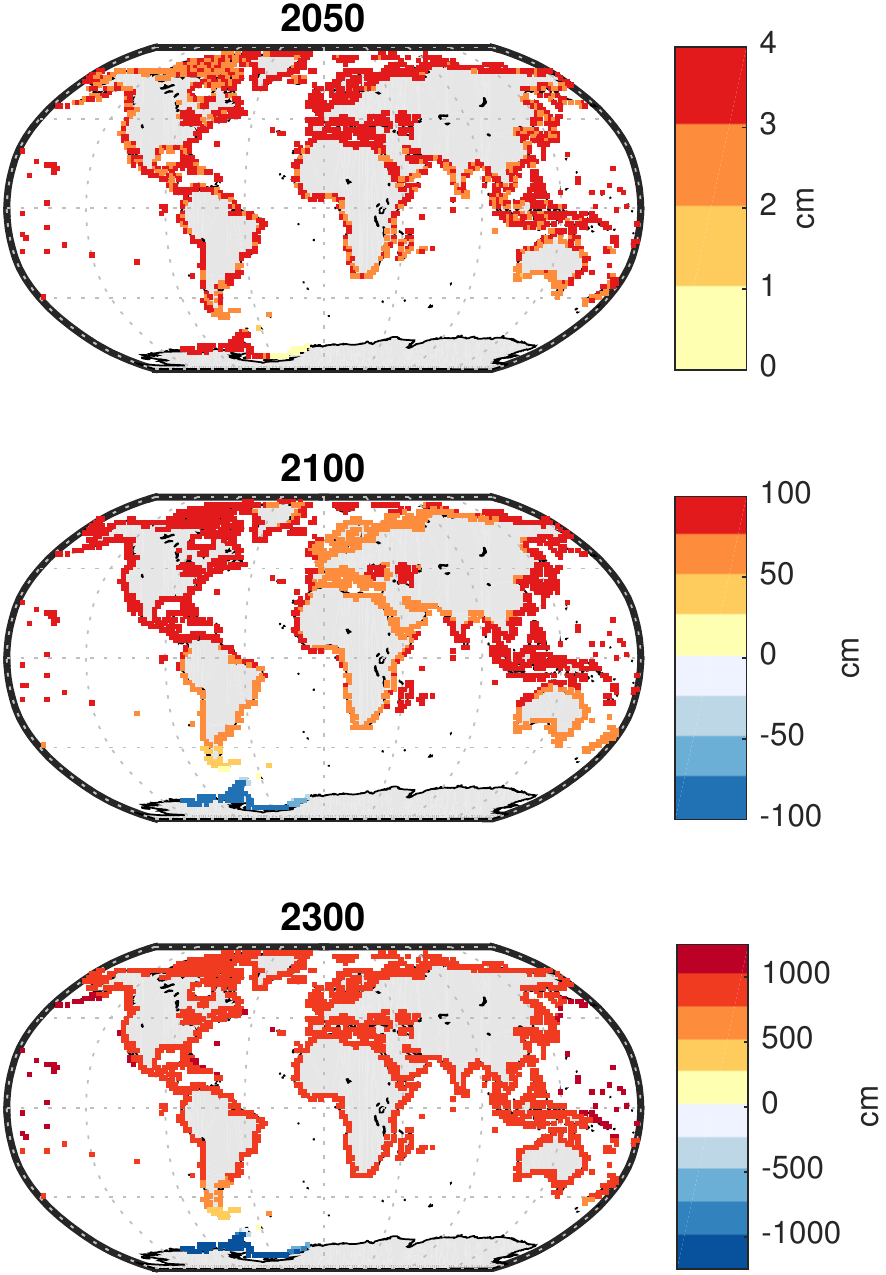}
\caption{Difference between DP16 and K14 projections in median RSL under RCP 8.5 in 2050, 2100 and 2300 .}
\label{Sfig:LSLproj_Pl_5-15minusK14}
\end{figure}

\begin{figure}[ht]
\centering
\includegraphics[width=20pc]{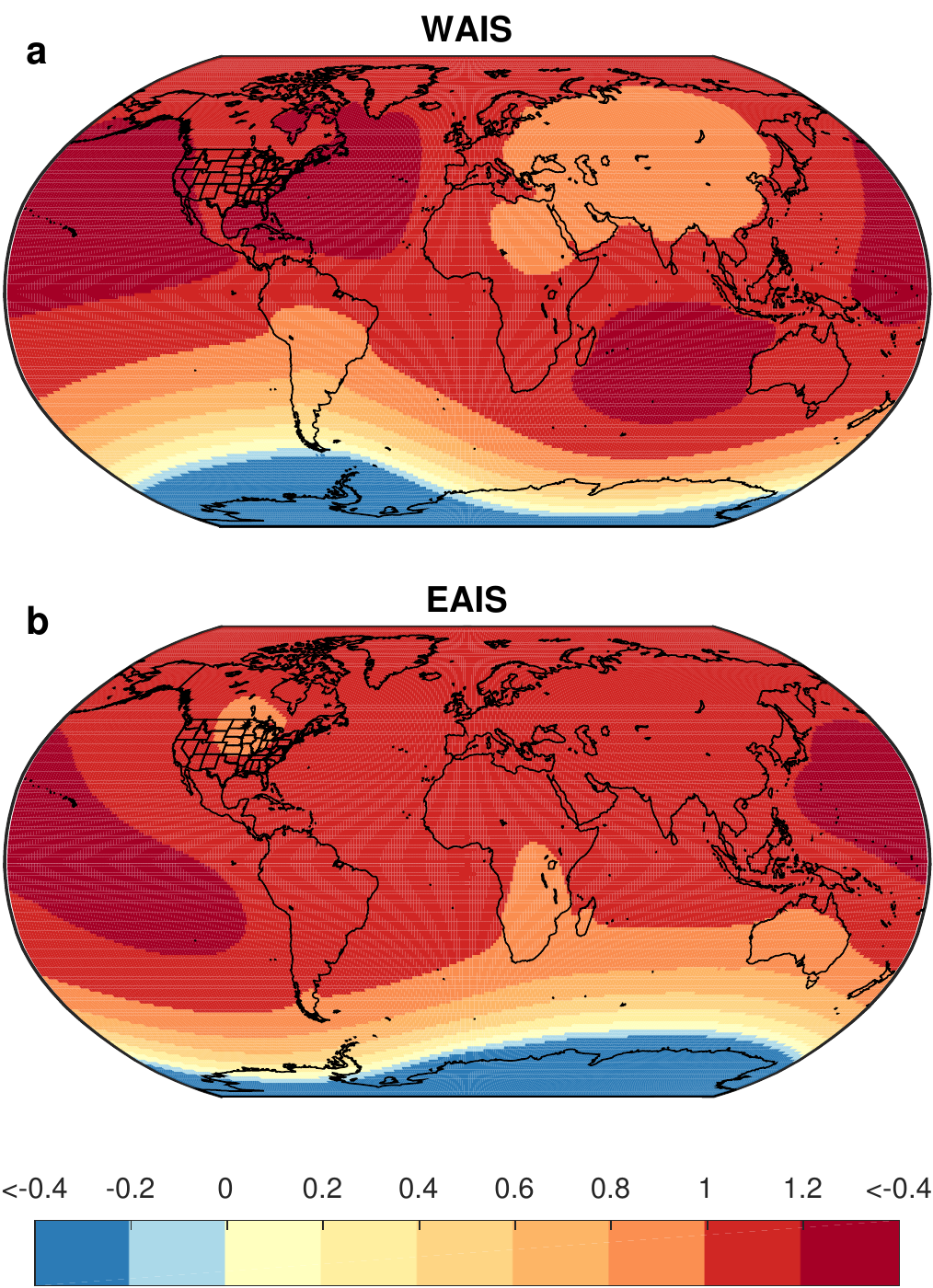}
\caption{Static-equilibrium relative sea-level fingerprints of (a) the West Antarctic Ice Sheet and (b) the East Antarctic Ice Sheet, as used in this analysis. Units are ratio of relative sea-level change to global mean sea-level change.}
\label{Sfig:fingerprints}
\end{figure}

\begin{figure}[ht]
\centering
\includegraphics[width=20pc]{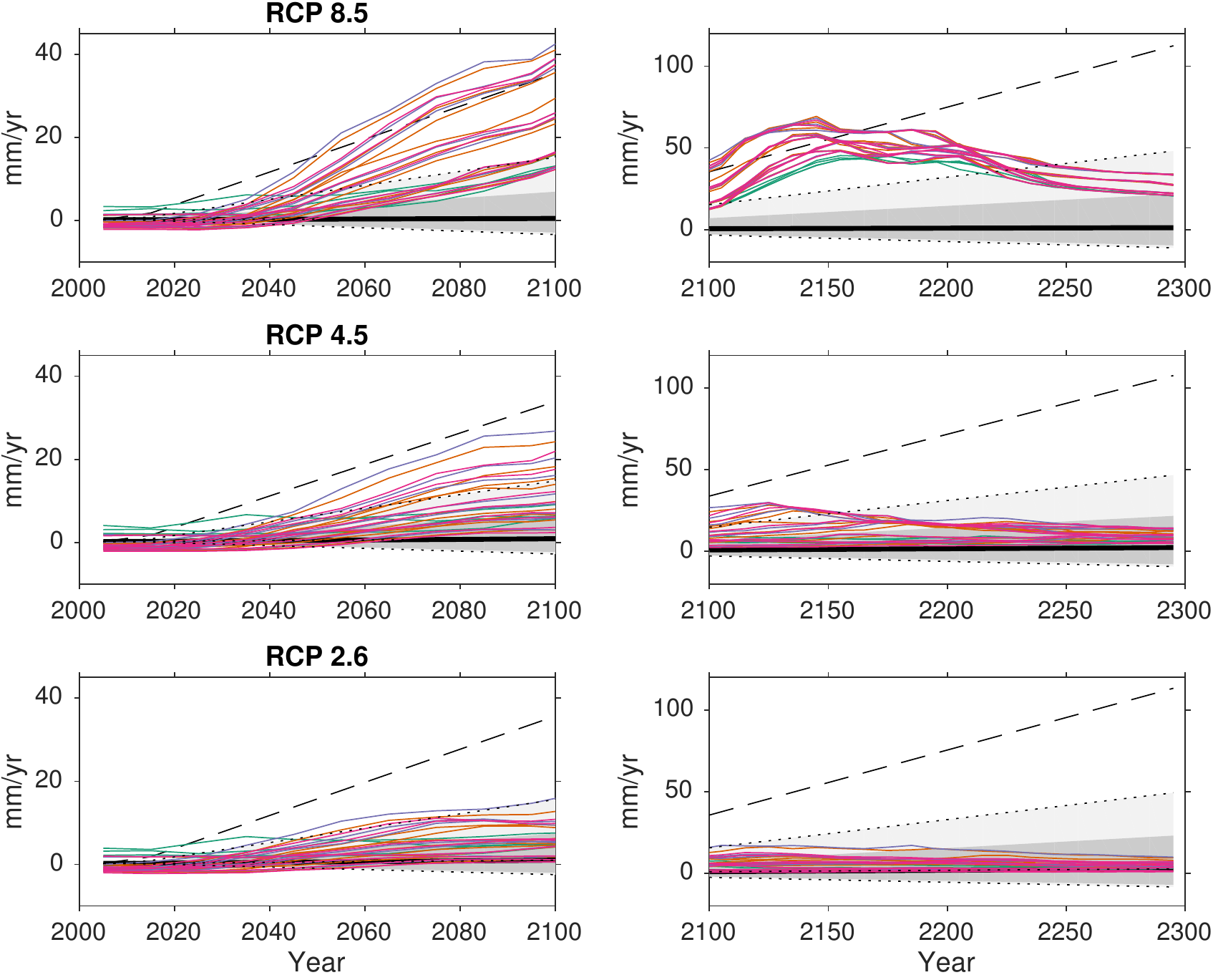}
\caption{Rates of contribution of the Antarctic ice sheet to GMSL under the three RCPs, with DP16 projections categorized by CREVLIQ. Dark/light shaded areas represent 5--95th and 0.5th--99.5th percentile of K14. Dashed black line represents 99.9th percentile of K14. Colored curves are DP16 runs (green: no ice cliff collapse; orange: 50 m per (m/yr)$^{-2}$; purple: 100 m per (m/yr)$^{-2}$; magenta: 150 m per (m/yr)$^{-2}$). Left panels show 2000--2100, right panels show 2100--2300. Note change of horizontal and vertical scales.  }
\end{figure}

\begin{figure}[ht]
\centering
\includegraphics[width=20pc]{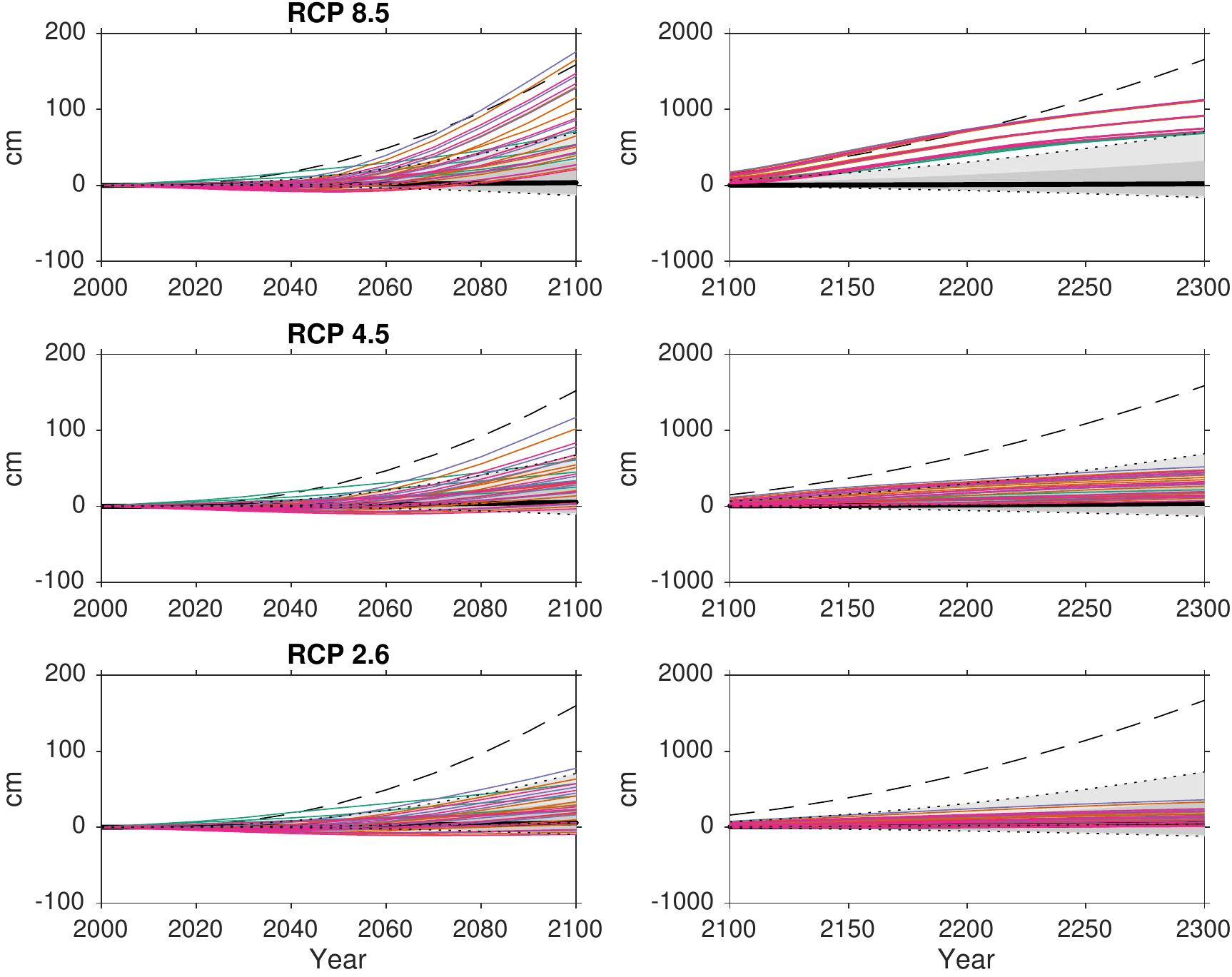}
\caption{Projections of the Antarctic ice sheet contribution to GMSL under the three RCPs, with DP16 projections categorized by CREVLIQ. Dark/light shaded areas represent 5--95th and 0.5th--99.5th percentile of K14. Dashed black line represents 99.9th percentile of K14. Colored curves are DP16 runs (green: no ice cliff collapse; orange: 50 m per (m/yr)$^{-2}$; purple: 100 m per (m/yr)$^{-2}$; magenta: 150 m per (m/yr)$^{-2}$). Left panels show 2000--2100, right panels show 2100--2300. Note change of horizontal and vertical scales. }
\label{Sfig:AISoldandnew-CREVLIQ}
\end{figure}

\end{document}